# Toward All 2D-based Printed Raindrop Triboelectric Nanogenerators

*Foad Ghasemi[1*], Jonas Heirich[1], Dimitri Sharikow[1], Sebastian Klenk[2], Jonathan N. Coleman[3], Georg S. Duesberg[2], Claudia Backes[1*]*

1. Physical Chemistry of Nanomaterials and CINSaT, Kassel University, Heinrich-Plett-Str. 40, 34132, Kassel, Germany.

2. Institute of Physics, University of the Bundeswehr Munich, Werner-Heisenberg-Weg 39 85577 Neubiberg, Germany.

3. School of Physics and CRANN, Trinity College, Dublin 2, Ireland

**ABSTRACT**

The raindrop triboelectric nanogenerator (RD-TENG) is an emerging technology that is designed to harvest energy from raindrops. This application requires materials with negative triboelectric effect, high surface charge density, mechanical flexibility, and a large surface area, which are key characteristics of 2D materials. However, fundamental research is necessary to understand the potential of 2D materials in this context. This study introduces all-2D-based RD-TENG devices using graphene and transition metal dichalcogenide (TMD) nanosheets. Liquid phase exfoliation (LPE) and liquid cascade centrifugation are used for nanosheet preparation and size selection. The TENGs are fabricated through a rapid, low-cost solution deposition technique based on liquid-liquid interface deposition, which allows screening of different active films and device geometries. Among the tested layered materials, medium-sized molybdenum disulfide ($MoS_2$) nanosheets (average lateral size~160 nm, volume-fraction weighted average layer number ~9) exhibit the highest short-circuit current (µA per drop) and voltage (mV per drop) output due to their most suited electron affinity, capacitance, and surface charge exchange properties. The variations in the performance of the TMD films were further evaluated with X-ray photoelectron spectroscopy (XPS), showing the influence of oxidation differences on charge transfer and charge decay time.

**INTRODUCTION**

Rapidly growing economy/population raises energy demands, making the development of alternative methods to generate green and renewable energy one of today's most important research questions. Although several technologies are already on the market, some possible energy sources remain unused, as technologies are still in the early development stages. Triboelectric nanogenerators (TENGs) are among these promising early development strategies[1]. The TENG converts mechanical motions into electrical energy through triboelectricity, a process in which materials gain or lose electrical charges when they come into contact[2]. The charge transfer occurs through contact electrification and electrostatic induction[3]. TENG devices are typically composed of metal contacts and triboelectric films, however, they can operate in various modes, including single-mode contact, double-mode contact, sliding, and triboelectric freestanding layers[4]. This versatility in geometry allows TENGs to be applied to a wide range of applications[4].

Recent advancements in TENGs have led to the design of raindrop-based TENGs (RD-TENGs), which were first described in 2014 and harvest energy from raindrops[5]. Unlike traditional TENGs, RD-TENGs rely significantly on surface properties to enhance performance[6]. Most currently reported RD-TENGs use polymer structures due to their simple fabrication process and flexibility[7]. However, the performance of these polymer-based devices is limited due to their low capacity for surface modification and surface charge density. This has led to the emergence of 2D materials as promising alternatives that can improve TENG's performance. Among the different 2D families, transition metal dichalcogenides (TMDs) are a suitable material platform, as they are well-established and demonstrate great potential for electronic devices[8]. Interestingly, TMDs exhibit various triboelectric tendencies and occupy different positions in the triboelectric series, which makes them particularly suitable for TENG applications[9]. Despite these promising advantages, TMD incorporation in RD-TENGs is little explored due to a lack of knowledge about their surface properties, in particular on interaction with water, as well as difficulties in achieving uniform, dense, large-area films on arbitrary substrates.

One promising approach to addressing these challenges is liquid-phase exfoliation (LPE), as it facilitates the scalable production of 2D flakes and subsequent deposition e.g. through printing, which is critical for RD-TENGs application[10]. To date, typical approaches to using LPE nanosheets to improve TENG performance are based on using LPE nanosheets as fillers in a polymer matrix. Here, a homogeneous distribution at a range of loading levels is crucial. Alternatively, one can envisage using 2D material-based thin films without a polymer matrix[11]. However, the lack of research on the surface properties of 2D films and challenges in engineering heterostructures that meet the geometrical requirements for RD-TENG devices, thus far has prevented systematic studies. The liquid-liquid interface deposition (LLID) technique, which is a modified Langmuir–Schäfer deposition, can address this limitation, as it can assemble 2D nanoflakes into uniform films with acceptable thickness control and surface coverage on arbitrary substrates over a few cm$^2$ areas on the labscale without the need for dedicated equipment[12]. Importantly, the tiled nanosheet networks produced this way have a significantly lower porosity and show better nanosheet alignment than other deposition methods[13, 14]. This implies that interfaces in heterostructures are relatively well defined which can be beneficial for RD-TENGs.

In this study, LLID deposition is utilized to create uniform films from LPE-exfoliated transition metal dichalcogenides, including tungsten disulfide ($WS_2$), molybdenum disulfide ($MoS_2$), tungsten diselenide ($WSe_2$), and molybdenum diselenide ($MoSe_2$) on conductive substrates suitable as RD-TENGs. The study investigates how flake sizes and deposition cycles relate to optimizing device efficiency and how surface properties such as wettability, surface coverage, and chemical composition impact the RD-TENG performance. To develop the assembly of all 2D-based RD-TENGs, LLID-deposited LPE-derived graphene fractions were analyzed for their thickness and electrical conductivity to identify the most promising network to serve as an electrical contact. The study provides a cornerstone for understanding material-structure-property relationships in 2D-based RD-TENGs, which might bring insight into their potential integration with other energy-harvesting systems, such as solar cells. Importantly, it shows that functional, simple devices can be fabricated using inexpensive, scalable production and deposition techniques.

**RESULTS AND DISCUSSIONS**

*Optimisation of network morphology and nanosheet size*

In this work, we focus on sonication-assisted liquid phase exfoliation of TMDs in a water surfactant following purification and size selection using cascade centrifugation[10, 15]. This is an iterative centrifugation process with sequentially increasing centrifugation speeds. After each step, sediments are collected, and the supernatant is subjected to the next step. Our typical sample nomenclature uses the lower and upper centrifugation boundary, e.g., 2.5-5 krpm. Unless otherwise noted, we use "medium-sized" nanosheets, which means that both large and thick, as well as small and thin nanosheets, were removed. After centrifugation, sediments are collected, and the solvent is exchanged to isopropanol (IPA) to facilitate deposition. Here LLID, a modified Langmuir-Schäfer deposition, is used[12]. In brief, the nanomaterial in IPA is injected at a liquid-liquid interface (water-hexane) until the interface is visually covered with the ink. The volatile hexane is allowed to evaporate, and the substrate is then lifted through the nanomaterial surface to transfer it onto the substrate. After drying, a homogeneous film is obtained. To build up stacks or increase the film thickness, the LLID deposition process can be repeated over the as-deposited film, even up to 15 deposition cycles called 15x.

Initially, a vertical RD-TENG with WS$_2$ sandwiched between bottom and top LPE graphene contacts was designed as a proof-of-concept for LPE-derived TMD films. The first device was composed of 5 deposition cycles of graphene film (5x) as a bottom contact, 1x of WS$_2$, and 5x of top graphene contact, but no detectable current/voltage was observed after the impact of tap water droplets. Subsequently, the WS$_2$ thickness was increased to 5x, resulting in weak but measurable signals. It was then suspected that high contact resistance was a limiting factor. To address this, the bottom and top graphene films were increased to 10x while 5x WS$_2$ film remained constant. This structure resulted in an improved performance, which confirms the importance of the contacts and charge separation in RD-TENGs (Figure S1). However, this fabrication process has a low yield and is time-consuming. Additionally, it caused increased surface roughness, which inhibited droplets from rolling off the surface, negatively impacting device efficiency.

To better understand the performance of 2D-based TENGs, the behavior of different sizes of one of the TMDs was first examined using simpler devices on ITO as electrodes with varying numbers of deposition cycles through the LLID deposition technique. The LPE and size selection methods are used here to prepare three WS$_2$ fractions with varying sizes/thicknesses. In Figure 1a, the normalized extinction spectra of the WS$_2$ fractions are shown, indicating an apparent difference between them. Extinction spectroscopy not only provides information about concentrations but also provides good information about nanosheet thickness and lateral size[16]. As a result, spectroscopic metrics for graphene and TMDs were developed that extract this information from an extinction spectrum[17]. According to the following extinction metric, the average length <L> and volume-fraction weighted mean layer number, <N>$_{vf}$ of the sheets can be approximately calculated for the WS$_2$ dispersion[16]:

$$\frac{ext(235)}{ext(295)} = \frac{0.0159<L>+2.20}{0.0166<L>+1} \quad (1)$$

$$\langle N \rangle_{Vf} = N_{bulk} \cdot e^{R(E_A - E_{A,bulk})} \qquad (2)$$

where ext(235) and ext(295) refer to extinction values at 235 and 295 nm, respectively. $E_A$ is the measured exciton energy with $E_{A,bulk}$, referring to the bulk material. R is a phenomenological material-dependent decay constant describing how the monolayer exciton energy approaches the bulk value as layer number increases. Values of R and $E_{A,bulk}$ are taken from ref [16]. For the 1.3-2.5 krpm fraction, the average flake size is 268.0 nm with 14.2 layers; the 2.5-5 krpm fraction has an average flake size of 152.2 nm and 9.0 layers; and the 5-12 krpm fraction shows an average flake size of 69.6 nm with 5.5 layers. Figure S2 compares the <L> and $\langle N \rangle_{Vf}$ of all three $WS_2$ fractions. Note that $\langle N \rangle_{Vf}$ summarized above is larger than the arithmetic mean <N>, as thicker nanosheets are also laterally larger in LPE[18].

Using the LLID method, the dispersions were first deposited onto quartz substrates in up to five sequential films[12]. Figure 1b presents the extinction spectra for the medium-sized films corresponding to each deposition layer. The inset provides an optical photograph of these films, demonstrating that as the number of deposition cycles increases, the films appear darker compared to the initial layers, resulting in progressively higher extinction. To improve the imaging resolution in the SEM, the films were deposited on ITO substrates. In Figure 1c, the films of large, medium, and small flake fractions are shown from left to right after one deposition cycle (1x). All three fractions demonstrate uniform surface coverage across the substrate. Figure S3 shows the SEM images of 1x, 3x, and 5x $WS_2$ films for all three fractions. The surface coverage for each fraction was determined from SEM images taken at the same magnification and plotted in Figure 1d. The supporting file and Figure S4 discuss the method used for this calculation. Among them, the 1.3-2.5 krpm fraction exhibits the lowest surface coverage, while the 2.5-5 krpm fraction achieves better coverage. Large flakes tend to create significant voids, reducing surface coverage, whereas smaller flakes can cover the surface more completely. Further, medium-sized flakes provide the best surface coverage compared with smaller flakes, which have less overlap at the edges. Overall, after three deposition cycles (3x), more than 90% of the surface is covered, while any additional deposition cycles contribute to less than a 5% increase in coverage. As each deposition cycle can also increase surface roughness, three deposition cycles offer a good compromise between coverage and disorder in the film.

Next, films were prepared on ITO electrodes to be tested as RD-TENGs. In this configuration, $WS_2$ nanosheets are deposited onto isolated ITO contacts as shown in the inset of Figure 1e. This kind of device serves as RD-TENG and was exposed to water droplets falling onto the $WS_2$ surface. The short-circuit currents for 1x to 5x deposition cycles of medium-sized films are shown under impact of tap water droplets released from a height of 52 cm. Based on the results, as the number of deposition cycles increases, the short-circuit current also increases, peaking at approximately 100 nA for the 3x film. However, beyond 3x deposition cycles, the short-circuit current decreases. This reduction is likely due to two factors: (1) In the 3x film, surface coverage exceeds 90%, significantly higher than in the 1x or 2x film, and further deposition cycles do not significantly change this coverage, and (2) additional deposition beyond 3x results in increased surface roughness and a certain open pore volume, which increases surface hydrophilicity and decreases the contact area between the surface and the impacting droplets. In principle, this effect can be tested with contact angle measurements. However, as discussed later, contact angle measurements for films beyond 3x were not feasible since the increased surface roughness led to higher standard deviations in the measured angles, making accurate assessment difficult. However, the effect can be observed when TENG device tests are performed, as the increased roughness prevents droplets from rolling off smoothly, which is critical for maintaining device performance and harvesting energy efficiently. As a result, both factors directly affect the final performance of the devices, and the 3x deposition is a good compromise between surface coverage and roughness.

Figure 1f compares the performance of 3x WS$_2$ films with small, medium, and large flakes on ITO contacts. These films were tested with tap water and 1 M NaCl droplets released from a height of 52 cm. The medium-sized WS$_2$ films generated the highest short-circuit currents (I$_{sc}$) of approximately 100 nA and 0.6 µA for tap and ionic water droplets, respectively. The larger signal observed with ionic droplets is attributed to their higher ion concentrations, which enhance charge transfer upon impact[19]. In contrast, the small and large flake films generate lower signals than medium-sized films. This difference is likely due to reduced surface coverage in the small flake films and increased surface roughness in the large flake films.

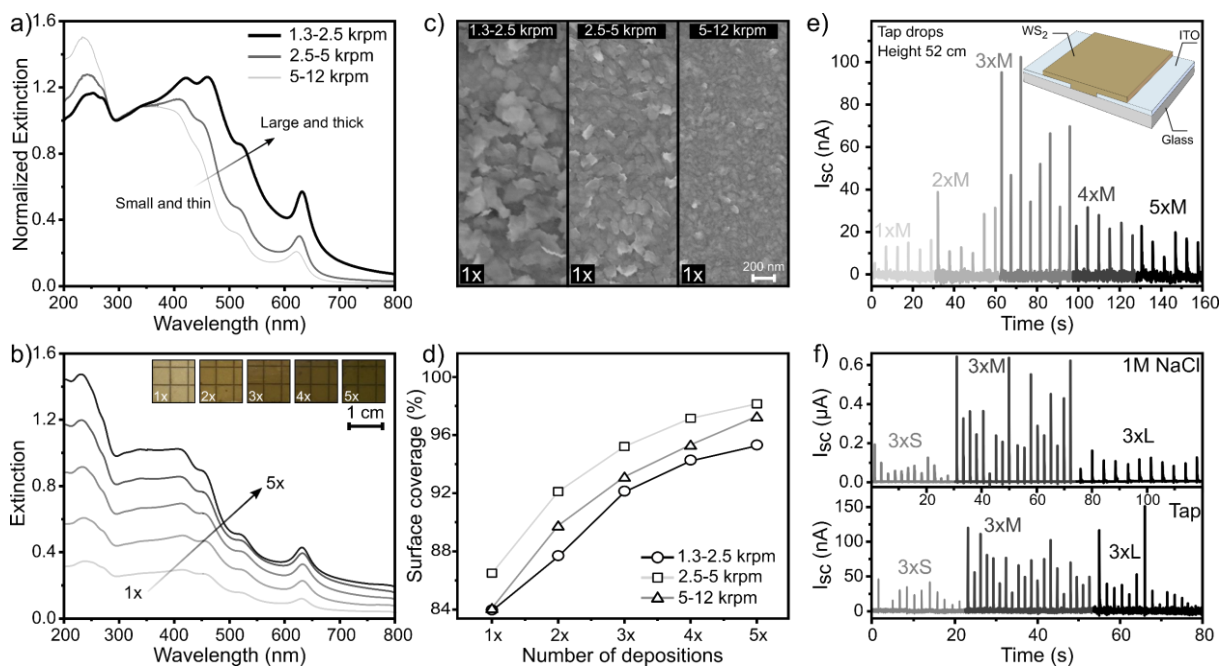

Figure 1. a) Normalized extinction spectra of WS$_2$ fractions with varying flake sizes and thicknesses. b) Extinction spectra of medium-sized WS$_2$ films after 1x to 5x LLID deposition cycles. Inset: optical photograph of the films. c) SEM images of large, medium, and small WS$_2$ flakes after 1x deposition. d) Surface coverage of WS$_2$ films for 1x to 5x deposition cycles of small, medium, and large flakes. e) Short-circuit current of medium-sized WS$_2$ films on ITO electrodes with 1x to 5x deposition cycles under the impact of tap water droplets. Inset: RD-

TENG device geometry on ITO/glass substrate. f) Short-circuit current comparison of 3x films of small, medium, and large $WS_2$ flakes on ITO electrodes under the impact of tap water and 1 M NaCl droplets.

*Material comparison*

To compare the performance of other TMD films, the same exfoliation process is used to prepare medium-sized $WS_2$, $WSe_2$, $MoS_2$, and $MoSe_2$ flakes with comparable average nanosheet lateral size and layer number. The extinction spectra of the prepared dispersion are plotted in Figure S5. Table S1 presents the size and thickness of the exfoliated TMDs derived from extinction spectra according to the spectroscopic metrics[16]. SEM images of the 3x deposited TMDs films are shown in Figure S6. The 3x films of all TMDs are deposited on ITO electrodes to serve as RD-TENGs. A series of devices were employed to evaluate the short-circuit current under releasing deionized (DI) water, tap water, and 1 M NaCl drops. Figure 2a illustrates the $I_{sc}$ response to DI water drops. All TMDs demonstrated relatively low $I_{sc}$ values for DI water, although $MoS_2$ and $WS_2$ showed a larger current change than other TMDs. Moreover, there were no distinct peaks for each impact of droplets, indicating relatively low sensitivity to water with low ionic conductivity. The $I_{sc}$ results for tap water droplets are shown in Figure 2b. In contrast to DI water, each TENG responded to tap water droplets with distinguishable signals. In tap water, ions improve conductivity and facilitate more effective triboelectric charge transfer upon impact, resulting in improved performance[19]. Figure 2c presents the $I_{sc}$ measurements for the 1 M NaCl solution, which contained the highest ionic concentration among the solutions. For all TMDs, clear and well-defined current spikes were observed, although their intensities varied. The results show that higher ionic strengths enhance the triboelectric effect since ions at the interface between a TMD and water improve charge transfer[19]. Accordingly, the $I_{sc}$ intensity increased progressively with increasing ionic strength from DI to tap water, then to 1 M NaCl drops across all TMDs.

Due to variations in the spike intensities for all TMDs, a series of devices (6 devices) were fabricated for each TMDs, and $I_{sc}$, the open circuit voltage ($V_{oc}$), and charge of all of them were measured under impact of 1 M NaCl droplets (78 drops per device). The results are summarized in the form of histograms to evaluate the overall performance. Figure 2d shows the maximum $I_{sc}$ spike intensities plotted in histograms for all TMDs. Each TMD has a relatively normal distribution showing specific central peak positions. The average $I_{sc}$ values of 0.09, 0.25, 1.11, and 3.33 µA were obtained for the $WSe_2$, $MoSe_2$, $WS_2$, and $MoS_2$ RD-TENGs, respectively. The $V_{oc}$ of all devices is also measured and plotted as histograms in Figure 2e, showing the same behavior as $I_{sc}$ with average values of 0.60, 1.03, 2.56, and 5.21 mV, respectively. As will be discussed later, the wider distribution observed in $WSe_2$ is likely attributed to oxidation states that form during drop releasing and electrical measurements. Based on $I_{sc}$ and $V_{oc}$ measurements, $MoS_2$ exhibits the highest values among the TMDs, followed by $WS_2$, $MoSe_2$, and $WSe_2$. This is further confirmed by the histograms of the separated charges, which can be extracted from the integration of the $I_{sc}$ spikes displayed in Figure 2f. Despite having a wide distribution, the histograms still follow the same trend as $I_{sc}$ and $V_{oc}$. Note that the highest currents using $MoS_2$ and tap water are in the µA range, with voltages in the mV range per drop. The observation aligns with some trends in triboelectric nanogenerators (TENGs), especially polymer-based raindrop TENGs[20, 21]. However, most polymer-based raindrop TENGs report higher open-circuit voltages while maintaining similar currents[22, 23]. It indicates that further surface or interface engineering of MoS2 to boost its output may be achievable.

$MoS_2$'s superior performance can be attributed to its highest electron affinity of ~4.3 eV which facilitates the build up of negative charge through interaction with water. $WSe_2$ exhibits the lowest $I_{sc}$ and $V_{oc}$ values, possibly due to its lowest electron affinity of ~3.8 eV. In total, device performance follows the order $MoS_2$ > $WS_2$ > $MoSe_2$ > $WSe_2$, which is roughly in the same order as the electron affinities of 4.3 eV ($MoS_2$), 4.1 eV ($WS_2$ and $MoSe_2$), 3.8 eV ($WSe_2$) according to ref [24]. Note that these values for electron affinities are theoretical values (DFT, bulk), as to date, no experimental values for this type of nanosheet network have been reported.

While the trend of electron affinities is well reflected in device performance, we note that other factors play a role, including the TMD films' wettability, surface roughness, and capacitance properties. Surface roughness and capacitance variations also affect charging and decay dynamics, influencing the overall performance and efficiency of RD-TENG materials[25]. In addition, upon the impact of water droplets, oxidation can degrade the electrical conductivity and stability of these TMD films, especially selenide-based films, which have a lower $I_{sc}$ and $V_{oc}$ signals with wider distributions.

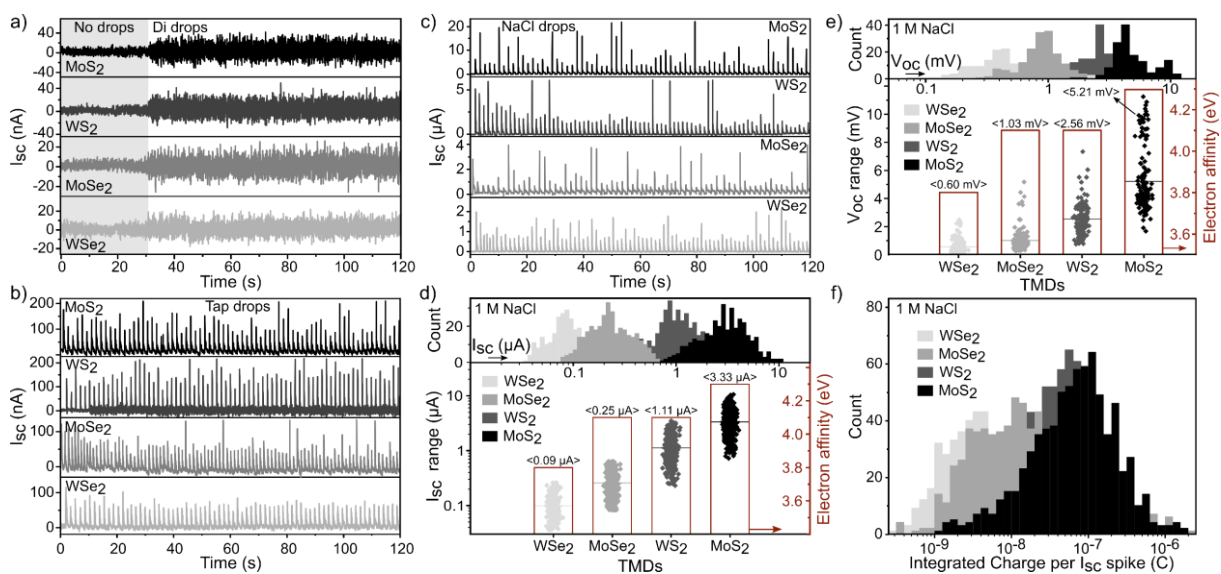

Figure 2. Short-circuit current ($I_{sc}$) response of TMD-based RD-TENGs ($MoS_2$, $WS_2$, $MoSe_2$, and $WSe_2$) to (a) DI water, (b) tap water, and (c) 1 M NaCl droplets. Histogram of all measured (d) $I_{sc}$ and (e) $V_{oc}$ spike intensities for a series of TMDs devices under 1 M NaCl drop releasing. (f) Charge histogram for all TMD devices derived from the $I_{sc}$ spikes.

To rationalize device performance, we conducted a series of additional experiments and data evaluation. Figure S7 shows the contact angle measurements of the TMD films on bare silicon substrates. Bare Si has a contact angle of $49.09 \pm 0.3°$. However, once the first film is deposited, the angle increases to a larger value for all TMDs. The second and third deposition cycles result in a further increase. This can be rationalized by the increasing surface coverage with each deposition cycle. At the same time, an increased surface roughness with each cycle increases the standard deviation in the contact angle measurement, as illustrated in the inset of Figure S7. Due to the larger standard deviation of the 3x films, the results are presented in histogram form. The histograms show that 3x $MoSe_2$ and $WSe_2$ exhibit contact angles of around 90°, while 3x $MoS_2$ and $WS_2$ have contact angles of around 80°. These results indicate that drops have more surface interactions on the $MoS_2$ and $WS_2$ films, which leads to more surface charge exchanges. This can be a factor contributing to the higher electrical short circuit currents in the sulfide-based films compared with the selenide-based films.

*Circuit model for device performance*

To further evaluate and understand the factors crucial for device performance, we make use of a circuit model presented below. Figure 3a illustrates a simplified equivalent circuit for a TMD/ITO RD-TENG. In this model, each TMD flake is considered as a capacitor named $C_{TMD}$ that is connected to other TMDs through a junction resistor named $R_j$. Consequently, the equivalent circuit model consists of several capacitors and resistors. As a result of equivalent capacitors and resistors, the entire device can be simplified into an RC circuit, where R and C represent the circuit's total resistors and capacitors.

This can be linked to the measured current, as each $I_{sc}$ spike decreases exponentially in electrical measurements, which can be modeled as $I_{sc} = Q/\tau \, e^{-t/\tau}$ where Q is the charge exchange from drop to device and $\tau$ is the decay time equal to RC (Figure 3b). It is possible to utilize the fitting function of an exponential decay in the form $y = y_0 + A_1 e^{-(x-x_0)/t_1}$, to fit each $I_{sc}$ spike, where $t_1$ ($\tau$) is the decay time in our introduced model (Figure 3c). However, due to the complex behavior of some spike decays, the fitting function may not always provide a good fit to the data points. In such cases, we focus only on the spikes with a successful fitting process and extract the decay function from those results. The results of the fitting functions and the decay times are shown in Figure 3d-h. The average decay times of 0.08, 0.12, 0.28, and 0.35 s were measured for $MoS_2$, $WS_2$, $MoSe_2$, and $WSe_2$, respectively. A shorter decay time suggests a smaller RC product. The dielectric constants of the four TMDs are likely within the same order of magnitude, suggesting that their total capacitances are comparable as well[26]. Thus, the key difference in their performance primarily arises from variations in total resistance. A higher total resistance indicates weaker interflake connectivity, which hinders efficient charge redistribution upon drop impact. As a result, electrostatic induction within the contact area is reduced, leading to diminished performance. As a result, the fast decay time of $MoS_2$ film is probably due to its higher conductivity and possibly better charge transfer. In contrast, the longer decay time of $WSe_2$ could be due to its lower conductivity and lower surface interaction. While a lower surface interaction can be inferred from the contact angle measurements, a lower conductivity of $WSe_2$ compared to $MoS_2$ cannot be easily rationalized for defect-free nanosheets with comparable film morphology. As we will demonstrate later, nanosheet oxidation is the most likely reason for this behavior.

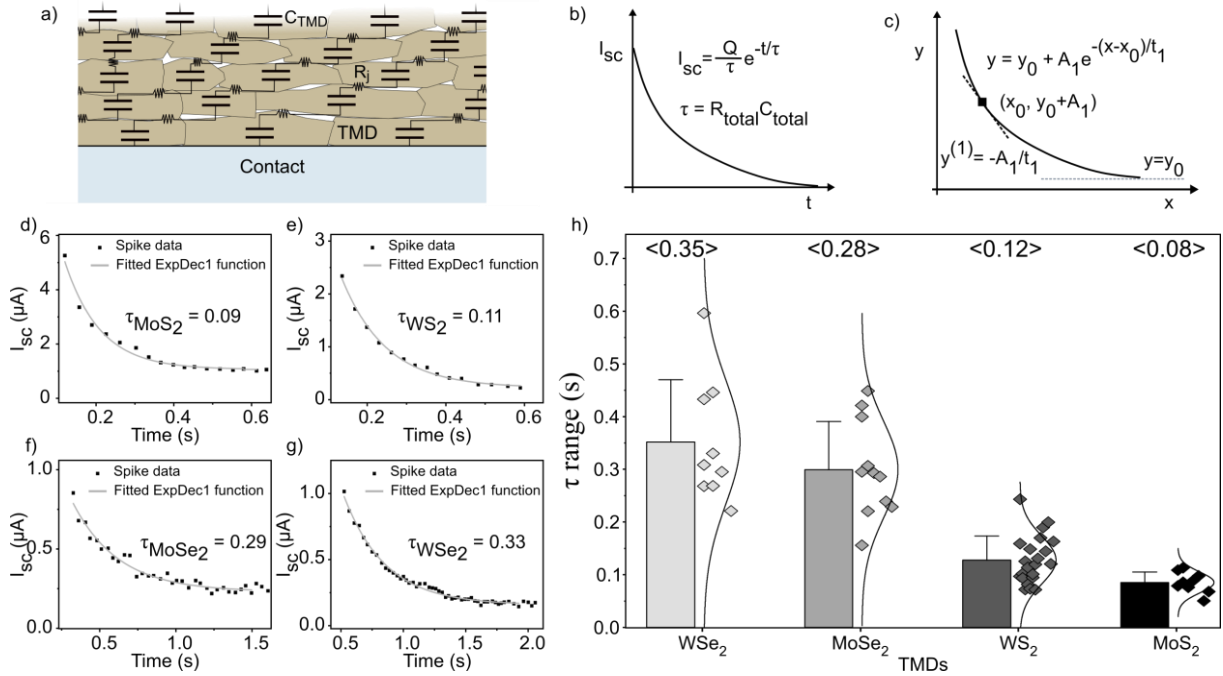

Figure 3. (a) Simplified equivalent circuit of TMD/ITO devices. (b) The introduced RC model for the $I_{sc}$ spike. (c) Decay fitting function used for the $I_{sc}$ spikes. Fitting functions and extracted decay times for one $I_{sc}$ spike of (d) $MoS_2$, (e) $WS_2$, (f) $MoSe_2$, and (g) $WSe_2$ RD-TENGs. h) Histogram plots of the extracted decay times for all four TMDs.

Note that this equivalent circuit model can also be applied to provide a better understanding of the flake size dependence discussed above. According to the capacitance equation, capacitance is inversely proportional to the thickness of a dielectric. To cover a fixed area, fewer large flakes are needed compared to small ones. As illustrated schematically in Figure S8, if each flake is considered as an individual capacitor, smaller flakes result in a greater number of capacitors connected in parallel. Since the total capacitance in a parallel configuration increases with the number of capacitors, films composed of smaller flakes exhibit higher overall capacitance. Based on size and thickness measurements of exfoliated $WS_2$ flakes, large flakes are approximately four times the size of small flakes. Consequently, covering a given area requires either one large flake, two medium flakes, or four small flakes, resulting in more parallel capacitors for films made with smaller flakes. Figure S8 also presents the equivalent circuit model for these films. Let $C_0$ represent the capacitance of a $WS_2$ monolayer, and N be the average number of layers per flake. Then, the capacitance of each flake is defined as $C_0/N$, with the following approximations: small flakes: $C_S = C_0/5.5$, medium flakes: $C_M = C_0/9$, and large flakes: $C_L = C_0/14.2$. Using this model, the total film capacitances are: $C^S_t \approx 0.48 C_0$, Medium flakes: $C^M_t \approx 0.14 C_0$, Large flakes: $C^L_t \approx 0.04 C_0$. This analysis shows that increasing flake size leads to a decrease in total capacitance, which correlates with reduced short-circuit current in films with larger flakes. Interestingly, although small flakes provide the highest capacitance, their poorer surface overlap and lower film density result in lower short-circuit current than medium flakes. Thus, the better short-circuit performance of medium-sized flakes is attributed to their optimal balance between surface coverage, surface roughness, and capacitance.

*Device stability*

To provide information on the stability of the devices, we performed extinction spectroscopy and XPS before and after exposure to NaCl and subsequent washing. For extinction spectra (Figure S9), 3x films of all TMDs were prepared on quartz substrates, and the extinction spectra were measured under three conditions: fresh films, after exposure to NaCl drops, and after washing with DI water. The extinction spectra of all films seem to remain similar to the fresh films in terms of spectral profile, except for variation in extinction intensity, which can arise from differences in the film's mounting in the spectrometer. The extinction did not drop systematically with the NaCl treatment and washing, so this cannot be attributed to the removal of the nanosheets on exposure to droplets. The exception is WSe$_2$ where the extinction spectrum of droplet-exposed films showed a relative increase in extinction in the UV region below 400 nm compared to freshly prepared WSe$_2$ films. This can be attributed to tungstates, i.e. a result of oxidation[27]. Interestingly, the extinction spectrum of WSe$_2$ returned close to the profile of the freshly prepared film after washing with DI water, albeit with significantly lower extinction.

This observation suggests that the water-soluble tungstates are removed through washing with water and do not remain in the film.

To validate this hypothesis, X-ray photoelectron spectroscopy (XPS) measurements were conducted on $WS_2$ and $WSe_2$ films to allow for a direct comparison of the metal core level spectra. The XPS analysis of the W 4f core levels in $WS_2$ and $WSe_2$ films reveals significant variations in chemical states before and after the release of NaCl drops. For the fresh $WS_2$ film (Figure 4a), the W 4f spectrum shows the characteristic W $4f_{7/2}$, W $4f_{5/2}$, and W $5p_{3/2}$ peaks at binding energies typical for $WS_2$, along with a second W 4f doublet at higher binding energies attributed to $WO_x$ compounds, probably due to surface oxidation states[28]. The atomic percentages confirm that the W 4f component of $WS_2$ is dominant (88.8%), with minor contributions from W 5p (6.5%) and $WO_x$ species (6.5%). After the release of NaCl drops (Figure 4b), the spectrum still shows the same components, but with an additional doublet at lower binding energies that constitutes ~3 at%. This suggests an electron transfer to the $WS_2$ due to the interaction between NaCl ions and the $WS_2$ film[29]. Further, the features assigned to $WO_x$ are reduced to 2.1 at%. We attribute this to a dissolution of the water-soluble tungstates by the treatment with the aqueous solution and, hence, removal from the film.

Similar features are observed for the $WSe_2$ film (Figure 4c), which includes W $4f_{7/2}$, W $4f_{5/2}$, W $5p_{3/2}$, and $WO_x$ peaks[30]. However, the $WO_x$ contribution is significantly higher (18.6%) than in the $WS_2$ film, which means a larger portion of the $WSe_2$ is oxidized already prior to treatment with the aqueous solution. After the release of NaCl drops (Figure 4d), a doublet at lower binding energy appears, which can be attributed to charge transfer species in the presence of NaCl, albeit with a lower at% than in $WS_2$ films. While this cannot be directly translated to the behaviour in TENG devices due to the absence of water, it nonetheless suggests a higher tendency to form charge-separated states in $WS_2$ compared to $WSe_2$, in agreement with the electron affinities discussed above. Similar to WS$_2$, the oxide content is also reduced in WSe$_2$ after exposure to aqueous NaCl, but much more significantly from 18.6 at% to 4.1 at%. Overall, a lower oxide percentage of WS$_2$ (5.6%) compared with WSe$_2$ (18.6%) suggests that LPE WS$_2$ is more resistant towards oxidation in the presence of water and oxygen. This improves charge interaction by reducing insulating barriers, resulting in better triboelectric performance. Furthermore, the interaction of NaCl ions with the cleaner WS$_2$ surface likely encourages stronger adsorption and charge redistribution. This results in improved electrical double-layer (EDL) formation and charge separation during triboelectric interactions. In contrast, higher surface oxidation leads to weaker ion interactions in the WSe$_2$ overall, resulting in a poorer triboelectric performance.

To confirm the hypothesis that water-soluble tungstates are removed through water and to test whether the charge separation in the presence of NaCl is reversible, the films, after exposure to aqueous NaCl, are washed with DI water, and XPS measurements are repeated. As shown in Figure S10, results reveal a significant reduction in the atomic percentage of oxides for both WS$_2$ and WSe$_2$ films. For the WS$_2$ film, the oxidation decreases to 1.6 at. Similarly, the WSe$_2$ film exhibits an oxide percentage of 2.3 at%, confirming the removal of the oxidized tungsten species. Further, after washing films with DI water, it seems that any residual ions resulting in charge-separated states are effectively also removed. This supports the hypothesis that ions bind weakly to the W interface, interacting through ionic or physical adsorption rather than forming strong covalent bonds. Further, it shows that the interaction with ions results in transfer of negative charge to the TMDs.

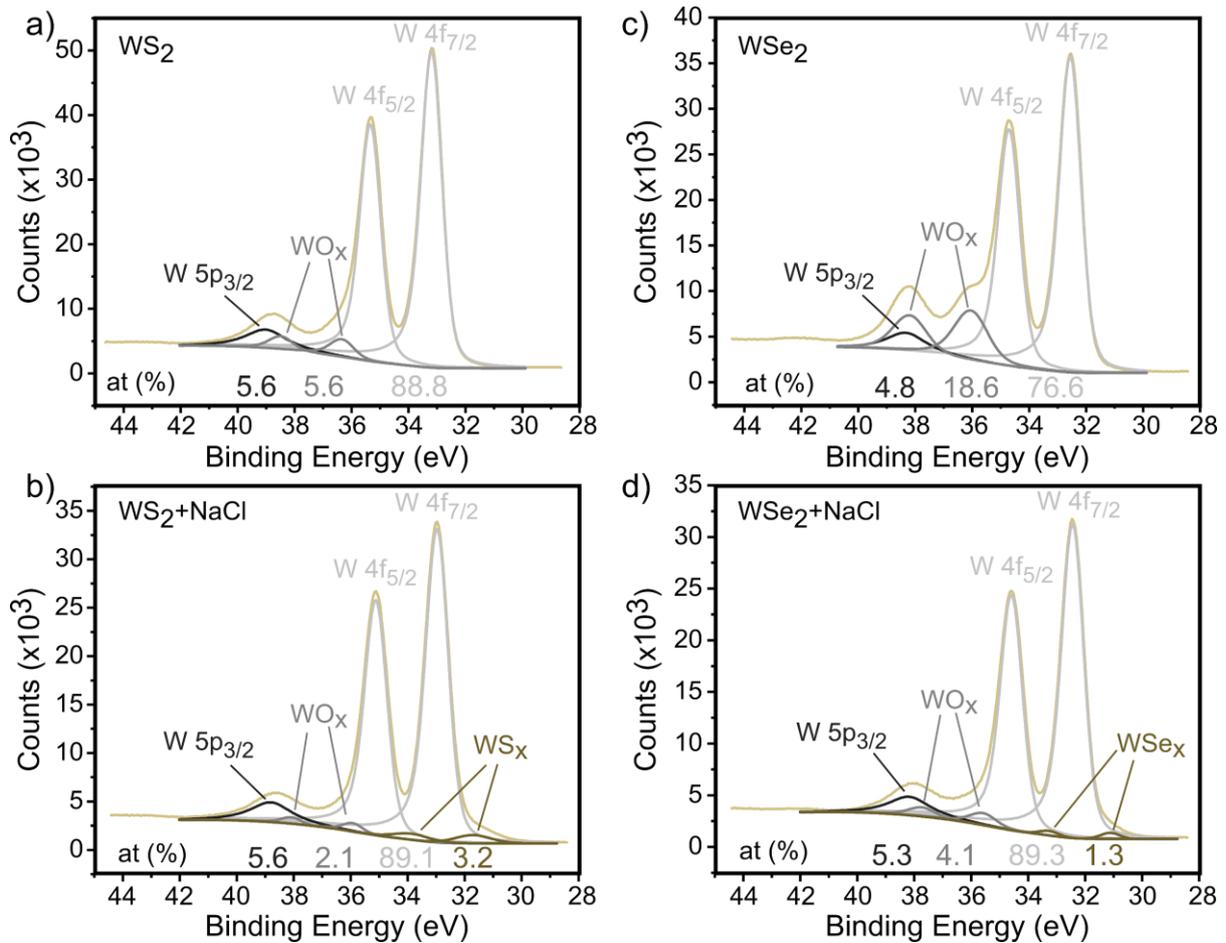

Figure 4. XPS analysis of the W 4f core in a-b) WS$_2$ and c-d) WSe$_2$ films before (a, c) and after (b, d) impact of NaCl droplets.

The XPS results align with the suggested response dynamics derived from the RC model. The model explains that the decay time ($\tau$) is proportional to the RC product, where a smaller $\tau$ indicates better charge transfer properties. By measuring the decay times of the $I_{sc}$ spikes, a shorter decay time was observed for sulfide-based films, while a longer decay time was obtained for selenide-based films. In agreement, XPS measurements reveal a lower portion of the electron rich species at lower binding in WSe$_2$ compared to WS$_2$ which translates to less participation of surface charges in short-circuit signals, a lower total capacitance and a longer decay time.

*Graphene electrodes*

To develop the RD-TENGs with all 2D materials, graphene is used as metal contacts. A similar LPE and size selection method was used to produce three graphene nanosheet dispersions with different sizes and thicknesses named 0.9-1.8 krpm, 1.8-2.9 krpm, and 2.9-6.8 krpm fractions denoted as large, medium, and small[17]. The goal of introducing these graphene fractions is to find films with good conductivity and straightforward deposition. Figure 5a displays the extinction spectra of graphene dispersions in sodium cholate (SC, 2 g/L) solution. By normalizing the spectra to the peak at ~271 nm (π–π* transition), a noticeable difference is observed due to variation in size and thickness[17]. It was experimentally reported and theoretically confirmed that the peak intensity ratio to the long wavelength plateau (~550–800 nm) increases with increasing nanosheet thickness[17]. The reported equations below were used to approximately estimate the arithmetic mean number of layers <N> and average nanosheet lengths <L> of graphene flakes[17]:

$$<N> = 35.7 \times \frac{Ext_{550}}{Ext_{325}} - 14.8 \qquad (3)$$

$$<L> = \left(\frac{Ext.exp}{2.466}\right)^{-3.05} \qquad (4)$$

where $Ext_{550}$ and $Ext_{325}$ are extinction values at 550 and 325 nm, respectively. An estimation of thickness is made using the extinction intensity ratio at 550 nm to 325 nm, and an estimation of the lateral size is determined by the exponent from fitting the extinction spectra between 550-800 nm with a power law. As depicted in Figure S11, the approximate values of <N> are 11.2, 8.4, and 5.5, with corresponding <L> values of around ~1036.6, 518.8, and 394.0 nm for the 0.9-1.8 krpm, 1.8-2.9 krpm, and 2.9-6.8 krpm graphene fractions, respectively.

Graphene dispersions were deposited onto quartz substrates with 1x to 15x deposition cycles using the LLID method (Figure S12a). As observed through visual inspection, the graphene films exhibit a lower opacity when initially deposited. However, as the number of deposition cycles increases, the films gradually become darker. Four-point probe measurements were carried out to evaluate the flake networks and film conductivities. Figure S12b displays the sheet resistance of graphene films with various deposition cycles from 1x to 15x. By increasing the number of deposition cycles, the conductivity of the films improves, which may be attributed to creating a more integrated three-dimensional network that enhances electrical connectivity. However, once the number of deposition cycles reaches approximately 10x, any additional deposition cycles do not significantly change the sheet resistance of the films. The conductivity of the initially deposited films is influenced by the area of overlap between the flake edges and by the voids between them. These factors affect the junction and network resistances of the film, respectively[31]. In our LLID-deposited LPE fractions, smaller and thinner graphene sheets exhibit higher sheet resistance compared to other fractions, likely due to their reduced overlapping area when deposited and more junctions per area. With additional deposition, there is a significant decrease in sheet resistance, attributed to the formation of a denser network. Table S2 lists the details of sheet resistance measurements for each film. Moreover, a comparison of medium and large graphene sheets shows that films composed of medium-sized sheets exhibit higher conductivity. This higher conductivity is likely attributed to fewer gaps between the flakes during the film formation in the deposition process. Although larger flakes may possess lower junction resistance, the increased voids between them can still lead to higher sheet resistance values.

A UV-visible spectrophotometer was used to evaluate the transparency of the graphene films. Figure 5b depicts the extinction spectrum of 1.8-2.9 krpm films with varying numbers of deposition cycles, ranging from 1x to 15x. As the number of depositions increases, the extinction values also increase, without any noticeable change in the π–π* peak except for detector saturation above extinction of ~3.5. Figure S12c plots the extinction value of all graphene films at 800 nm, which increases with increasing deposition cycles. The 0.9-1.8 krpm fraction exhibits the highest extinction values, whereas the 2.9-6.8 krpm fraction exhibits the lowest values for the same number of deposition cycles, in line with the corresponding flake thicknesses of the nanosheets in dispersion. This observation will become more significant in future work aimed at designing devices for harvesting both rain and sunshine simultaneously[32].

Figure 5c presents the relationship between sheet resistance and extinction values for all graphene films, making it easier to choose the most appropriate graphene film for RD-TENG devices. Here, graphene films with medium-sized flakes appear to be the best choice as they demonstrate higher transparency and higher conductivity, which makes them potentially more suitable for RD-TENG contacts compared to the other fractions. However, they still exhibit high sheet resistance, making them not feasible as an electrical contact in the current form. For example, the sheet resistance of 15x medium-sized graphene is around 327 $\Omega/\square$. To obtain this relatively suitable conductivity, a time-consuming deposition process must be undertaken. As mentioned above, the better performance of 10x graphene-based RD-TENGs compared with 5x devices in our initial tests highlights the significant role of electrode resistance. Although 10x graphene films exhibit a higher short-circuit current (Figure S1), the current is still much lower than that of ITO devices. This demonstrates that contact conductivities play an essential role in the device's performance. Overall, the number of depositions should be optimized to achieve the best possible performance. At first glance, more deposition cycles may seem desirable based on electrode conductivity, but this comes with some drawbacks, such as longer fabrication times, rougher surfaces, and reduced hydrophobicity, which may diminish the final RD-TENG performance.

Carbon nanotubes (CNTs) have been shown to significantly enhance the conductivity of graphene films[33]. As a result, single-walled carbon nanotubes (SWCNTs) were prepared using the LPE method in an IPA/DI mixture and added to the medium-sized graphene dispersions in various mass ratios. The details of the mass ratio calculations for graphene/SWCNT dispersions are discussed in the SI. These mixtures were then deposited on substrates using the LLID method with a hexane/DI interface. SEM images of the 5x Graphene films with 0, 27, and 75 wt% SWCNTs are shown in Figure 5d. By increasing the SWCNT content, the density of graphene flakes is decreased, and distinct graphene nanosheets are hard to discern in the 75 wt% SWCNT film.

Figure 5e illustrates the sheet resistance results for graphene and its composites with a SWCNT ratio of 0, 8 wt%, 27 wt%, 60 wt%, and 75 wt%. As summarized in Table S3, by increasing the SWCNT content, the sheet resistance decreases, and films show higher transparency. Compared with pure 3x graphene film, the 3x graphene film with 27% CNT content shows a 83% reduction in sheet resistance, from 3172.4 to 551.9 $\Omega/\square$, along with a 4% improvement in transparency (extinction value at 800 nm reduced from 0.508 to 0.487), making this combination a promising candidate for TENG electrical contacts. The sheet resistance and transparency are further improved with increasing the SWCNT content. However, contact angle measurements reveal that higher SWCNT contents increase the hydrophilicity of the films (Figure S13). The contact angle for 1x Gr film is measured to be 72.0°±2.29 on quartz substrates and it is decreased to 67.0°±2.56, 59.2°±2.06, 57.4°±1.77, and 56.6°±0.94 after addition of 8, 27, 60, and 75 wt% SWCNT, respectively. This decreasing contact angle with increasing SWCNT content is not desirable for RD-TENGs because droplets will remain on hydrophilic contacts and decrease the device performance. The hydrophilicity of the SWCNTs is attributed to the raw material which is pre-functionalized with carboxylic acid functional to facilitate processing.

Overall, a balance between contact angle, surface coverage, and sheet resistance must be maintained. Based on this, the 3x graphene film with 27 wt% SWCNT was selected as RD-TENG contacts with over 90% surface coverage and a sheet resistance of 551.9 Ω/□ (Figure 5d,), comparable to that of 8x to 10x films of pure graphene, while providing the additional benefits of reduced surface roughness, faster fabrication, and higher reproducibility.

To characterize the electrodes in more detail, the thickness of a range of films was measured using atomic force microscopy (AFM). The extinction values of graphene films should be directly correlated with their thickness in homogeneous tiled nanosheet networks. To confirm this, for medium-sized graphene, the film thicknesses were measured and plotted versus the extinction at 800 nm in Figure 5f. A linear relationship is observed between thickness and extinction with a slope of ~60. Importantly, the data agree well with a fit through the origin, which is expected, as zero thickness should result in zero extinction. The inset of Figure 5f also shows the data for Gr:CNT films containing 27 wt% SWCNTs. Here, fitting a linear function results in $t_{Gr:CNT}=46.02 Ext_{800}$ . This is a manifestation of Figure 5e that adding SWCNTs results in changing the transparency due to the distinct extinction spectrum of this type of SWCNTs (Figure S14).

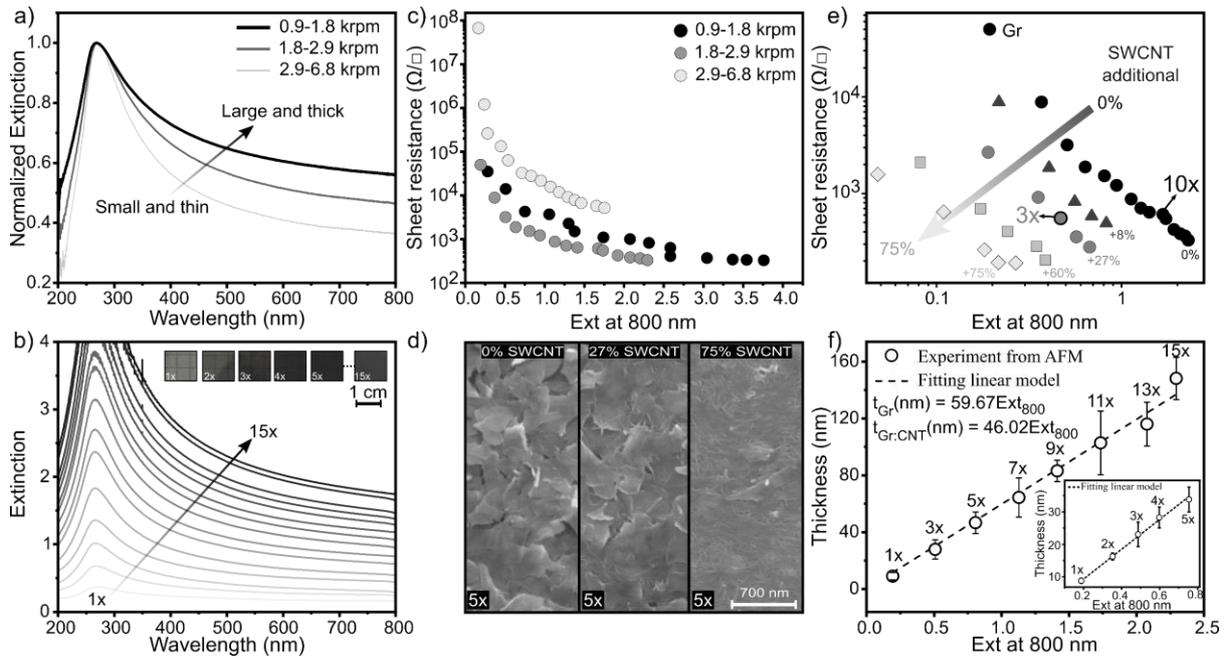

Figure 5. (a) Normalized extinction spectra of graphene dispersions, (b) Extinction spectra of the medium-sized graphene for 1x up to 15x deposition cycles. The inset shows the optical photograph of the films. (c) Sheet resistance vs. extinction values for graphene films. (d) SEM images of the 5x medium-sized graphene films with 0, 27, 60 and 75 wt% SWCNT content. (e) Sheet resistance vs. extinction values for Gr:CNTs films. (f) Thickness versus extinction value at 800 nm wavelength for Gr films. The inset shows the results for Gr:CNTs films with 27 wt% SWCNT ratio.

*All printed, all 2D-based devices*

Taking advantage of the improvement in electrical conductivity, RD-TENGs were designed using Gr:CNTs contacts and TMDs as the primary active materials. To enable precise patterning, custom molds and masks were designed and fabricated using 3D printing, which provided shadow masks for the solution deposition process using the LLID method. As shown in Figure 6a, glass substrates were placed within a printed mold, followed by an interdigitated mask attached to the mold. By tightening the screws between the glass substrate and mask, secure contact was ensured. Once assembled, the whole molds were placed in the deposition holder, where Gr:CNTs dispersions were deposited sequentially three times. An optical image of the deposited film is shown in the inset of Figure 6a. Following the deposition of Gr:CNTs, the interdigitated mask was replaced with a secondary mask for the deposition of TMDs. Over the Gr:CNTs electrode lines, TMDs were sequentially deposited by repeating the LLID for three cycles. In the inset of Figure 6a, the final device configuration can be seen. To enhance adhesion properties and optimize device performance, the assembled devices were vacuum-annealed at 150 °C for 2 hours after the final assembly. The $I_{sc}$ responses of TMD-based devices to DI water, tap water, and 1 M NaCl drops are shown in Figure 6b. When DI water droplets are released, $MoS_2$ shows the largest response, whereas $WSe_2$ shows the lowest signal. The same behavior is also observed for tap water and NaCl drops. This is consistent with results from TMDs/ITO TENGs, indicating that $MoS_2$ and $WS_2$ perform better than selenide-based devices, also in all solution-processed devices.

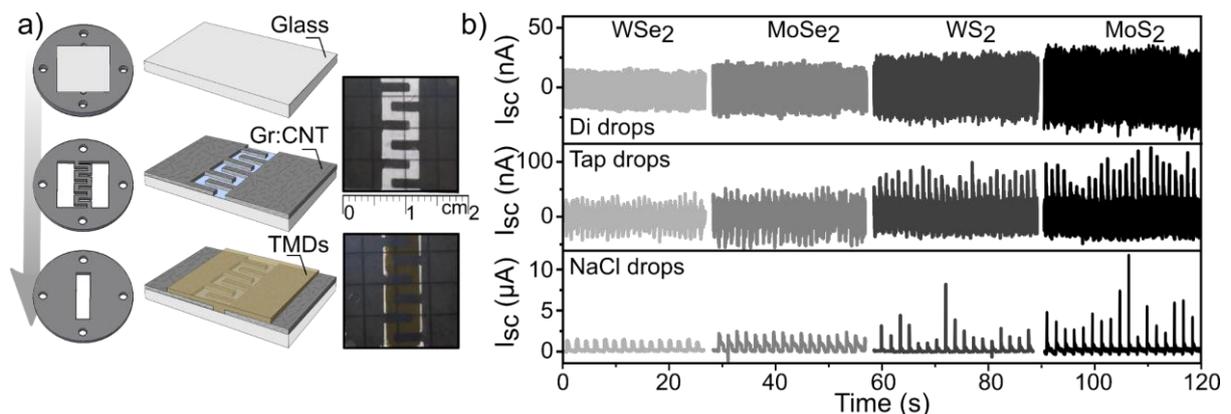

Figure 6. (a) The fabrication process for TMD-based raindrop TENGs utilizing Gr:CNTs interdigitated contacts and TMDs as active materials. The inset shows the photograph of the deposited films. (b) Short-circuit current ($I_{sc}$) responses of TMD-based devices to DI water, tap water, and 1 M NaCl drops.

By employing the same LLID method as used for the interdigitated electrode design, vertically structured 2D-based RD-TENGs are also introduced here. In this configuration, 3x films of Gr:CNTs are first deposited on a glass substrate to create the bottom contact. Afterward, 3x films of TMDs are deposited over two-thirds of the Gr:CNTs film, leaving one-third exposed. Using a second mask, another 3x films of Gr:CNTs are deposited on the remaining one-third area of the TMD film to form the top contact. The fabrication steps of the devices are schematically shown in Figure S15a. The $I_{sc}$ results for DI water, tap water, and 1 M NaCl drops are presented in Figure S15b, respectively. As observed in the interdigitated configuration, $I_{sc}$ increases with higher ionic concentrations, confirming enhanced charge transfer with increased ion presence. The vertical RD-TENG shows a higher $I_{sc}$ than the interdigitated RD-TENG for 1 M NaCl drops. This may be attributed to the optimized contact area and the enhanced charge separation efficiency in the vertical structure, which promotes efficient charge transfer across the TMD and Gr:CNTs interfaces[34]. The high ionic strength of the NaCl solution further enhances electrostatic interactions at these interfaces, facilitating greater triboelectric charge generation and a stronger current response[19]. This effect is less pronounced in DI water and tap water, which have lower ion concentrations and, therefore, do not exhibit the same degree of electrostatic enhancement as observed in 1 M NaCl. In the same way as previous devices, $MoS_2$ and $WS_2$ devices perform better than $MoSe_2$ and $WSe_2$ devices.

**CONCLUSION**

In this study, graphene and transition metal dichalcogenide (TMD) nanosheets are employed as all-2D-based raindrop TENGs. The films were prepared using LPE and LLID. Various sizes and thicknesses of the exfoliated TMD flakes were examined as the main triboelectric films of the devices. Among these, medium-sized $MoS_2$ nanosheets demonstrate the highest current and voltage output due to their better triboelectric tendencies and surface properties. Moreover, the circuit model and X-ray photoelectron spectroscopy (XPS) analysis illustrate the essential role of surface oxidation in determining the variations in charge transfer and charge decay dynamics across different TMDs. The ease of fabrication has great potential to investigate the TENG performance of other layered materials, while further optimization in terms of transparency might allow for an integration with solar cells.

**EXPERIMENTAL DETAILS**

**Materials.** Graphite powder (~325 mesh, 99% carbon basis), molybdenum disulfide powder (<2 μm, 99%), tungsten disulfide powder (<2 μm, 99%), and sodium cholate hydrate (≥99%) derived from bovine and/or ovine bile were obtained from Merck Sigma-Aldrich. Molybdenum diselenide (325 mesh powder, 99.9%) and tungsten diselenide (10–20 μm powder, 99.8%) were purchased from Alfa Aesar. A single-wall carbon nanotube powder (P3-SWNT, Purity > 90%) was provided by Carbon Solutions, Inc. The thin films were deposited on quartz substrates purchased from Precision Glass & Optics GmbH.

**Liquid phase exfoliation.** For TMDs, 20 g/L of the initial TMD powder was added to an SC solution with a 6 g/L concentration. The mixture was then sonicated for 1 hour at 55% amplitude with a 6:4 s on:off pulse sequence. After centrifuging at 4700 rpm for 1.5 hours, the supernatant was discarded, and the sediment was re-dispersed in 2 g/L SC solution. A 55% amplitude horn tip sonication was repeated for 5 h at the same pulse rate on the dispersion. Then, it was centrifuged at 1.3 krpm for 2 hours, and the sediment (unexfoliated flakes) was removed. The supernatants were centrifuged at 2.5 krpm, 5 krpm, and 12 krpm, and after each step, the sediments were collected in 0.1 g/L SC solution. These collected sediments were labeled as 1.3-2.5 krpm, 2.5-5 krpm, and 5-12 krpm fractions, respectively.

To prepare the graphene fractions, graphite powder was added to an aqueous SC solution (8 g/L in DI water) at a concentration of 30 g/L and sonicated with horn probe sonication (Sonics VXC-500) for 1 hour at 60% amplitude with a pulse rate of 6 s/2 s on/off. Then, the mixture was centrifuged at 5.9 krpm for 2 h, and the sediment was collected in 80 mL of a fresh 2 g/L aqueous SC solution. The dispersion was sonicated at 60% amplitude with the same pulse sequence for 5 h. Next, it was centrifuged at 0.9 krpm for 2h, the supernatant was collected, and the sediment (unexfoliated flakes) was discarded. Following centrifugation at 1.8 krpm,

2.9 krpm, and 6.8 krpm for 2 h, the sediments of each step were collected in fresh SC solution (2 g/L) labeled as 0.9-1.8 krpm, 1.8-2.9 krpm, and 2.9-6.8 krpm graphene fractions.

To exfoliate carbon nanotubes (SWCNTs), 0.5 g/L of P3-SWNT powder was dispersed in a 60% isopropyl alcohol (IPA)/40% deionized water (DI) mixture. The dispersion was sonicated with a microtip at 25% amplitude with a pulse sequence of 6 s on and 2 s off for 3 h. The mixture was centrifuged at 5 krpm for 2 h to separate larger agglomerates. The top supernatant, containing the well-dispersed CNTs, was carefully collected as the final CNT dispersion.

**Graphene/SWCNTs dispersion.** First, the medium-sized graphene dispersion in SC was diluted in DI and centrifuged at 15 krpm for 2 hours. Then, the sediments were collected and dispersed in an IPA/DI (60%/40%) mixture to allow straight forward mixing with the SWCNT dispersion. The concentration of the graphene and SWCNT dispersions in IPA/DI was calculated by measuring the UV-Vis extinctions and using the Beer-Lambert law. The extinction coefficient of the P3-SWCNTs was determined as 3292 mL.mg$^{-1}$.m$^{-1}$ through gravimetry after microfiltration onto alumina membranes. Then, the mass of flakes/tubes per volume was calculated, and different volumes of SWCNT dispersion were added to graphene dispersion to have mixtures with SWCNT mass ratios of 8%, 27%, 60%, and 75%.

**Solvent exchange of the 2D ink.** To transfer the 2D inks from SC solvents to IPA, the inks were first diluted in DI water and centrifuged at 15,000 rpm for 2 hours. The resulting sediments were collected and re-dispersed in IPA at varying concentrations, depending on the fractions. Specifically, concentrations of 1 g/L, 0.5 g/L, and 0.1 g/L in IPA were used for the large, medium, and small fractions, respectively. Prior to deposition, all inks were gently bath-sonicated for a few seconds to ensure uniform dispersion. In the case of graphene/SWCNT ink, the dispersion in IPA/DI was directly used for the deposition process.

**Liquid-liquid interface deposition.** The LLID experiment involved placing a substrate on a teflon stand in a beaker of deionized water. Over the submerged substrate, distilled hexane (1 mL) was added to form a water/hexane interface. Then, TMDs inks in IPA (or graphene/SWCNT in IPA/DI) were added inside the hexane until a continuous layer was created over the water/hexane interface. In the case of graphene inks, the dispersion was added directly to DI water to form a water/air interface. To coat the substrate with nanosheets, the teflon stand was slowly lifted through the formed layer. Once the films were dried, they were annealed in air at 60 °C for 15 min to enhance solvent removal and film adhesion. The next film deposition is done by repeating this procedure.

**Raindrop TENG Device Fabrication. TMDs/ITO Device:** For the TMDs/ITO device, the ITO/glass substrates were initially scratched to create two isolated regions, effectively forming source and drain electrodes. Then, TMD films were deposited on the whole substrate in varying numbers. **Gr:CNT/WS$_2$/Gr:CNT Device:** 3x Gr:CNT (or Gr) film was deposited onto glass substrates as a bottom contact. By using a 3D-printed mask, 3x TMDs film was deposited over the selected region of the previous films. The top contact was then coated with a 3x Gr:CNT (or Gr) film using a second 3D-printed mask to block portions of the TMDs film. **Interdigitated Gr/TMDs Device:** The glass substrate was placed inside a 3D-printed mold, and the interdigitated mask was attached on top of it. 3x Gr:CNT films were deposited through interdigitated patterns in a LLID deposition setup. The interdigitated mask was then replaced with another rectangular mask, and 3x TMD films were deposited over pairs of Gr:CNT interdigitated electrodes.

All devices were annealed at 150 °C in a vacuum for 2 hours to enhance film adhesion. Then, copper tapes were attached to the device surfaces. A PVC tape was then applied over the copper surface to ensure connection durability during exposure to water droplets.

**Raindrop measurements.** To simulate raindrop impacts on the devices, a Legato® 100 Syringe Pump from KD Scientific was used with different types of water, including deionized water, tap water, and 1 M NaCl solution at a controlled flow rate of 200 μL/min, releasing from a distance of 52 cm. Under simulated raindrops, the devices were positioned at a 45° angle. Keithley 2400 SourceMeter unit and Tektronix TDS2012 were used to measure short-circuit current and open-circuit voltage, respectively. A MATLAB program was implemented to collect real-time data.

**Characterization.** The extinction spectra of the dispersion were measured using an Agilent Cary 60 UV-Vis/NIR spectrophotometer through quartz cuvettes with a 1 cm path length. A Cary 5000 UV-Vis-NIR spectrophotometer was employed to measure the extinction spectra of thin films. All spectra were acquired in 0.5 nm increments. The sheet resistance of thin films was determined with an Alessi four-point probe connected to a Keithley 2400 SourceMeter unit. The surface morphology of the films was investigated using a HITACHI S-4000 scanning electron microscope (SEM) at an acceleration voltage of 10 kV. Atomic force microscopy (AFM) images were taken using a FlexAFM by Nanosurf in tapping mode under ambient conditions with a silicon cantilever (Nanosensors, type NCLR-W). Contact angle measurements were performed with a KSV's CAM 100 instrument. A drop of water was placed on the surface and recorded with the camera. XPS measurements were carried out using a PHI VersaProbe III equipped with a micro-focused monochromatic (Al Kα source, 1486.6 eV) and dual-beam charge neutralization. Binding energies were referenced to the carbon signal at 284.8 eV, and peaks were fitted using CasaXPS software.

## REFERENCES

(1) Du, T.; Chen, Z.; Dong, F.; Cai, H.; Zou, Y.; Zhang, Y.; Sun, P.; Xu, M. Advances in Green Triboelectric Nanogenerators. *Adv. Funct. Mater.* **2024**, *34* (24), 2313794.

(2) Wu, H.; Shan, C.; Fu, S.; Li, K.; Wang, J.; Xu, S.; Li, G.; Zhao, Q.; Guo, H.; Hu, C. Efficient energy conversion mechanism and energy storage strategy for triboelectric nanogenerators. *Nat. Commun.* **2024**, *15* (1), 6558.
(3) Jiang, Y.; Wu, Y.; Xu, G.; Wang, S.; Mei, T.; Liu, N.; Wang, T.; Wang, Y.; Xiao, K. Charges Transfer in Interfaces for Energy Generating. *Small Methods* **2024**, *8* (4), 2300261.
(4) Kim, W.-G.; Kim, D.-W.; Tcho, I.-W.; Kim, J.-K.; Kim, M.-S.; Choi, Y.-K. Triboelectric Nanogenerator: Structure, Mechanism, and Applications. *ACS Nano* **2021**, *15* (1), 258-287.
(5) Manojkumar, K.; Muthuramalingam, M.; Sateesh, D.; Hajra, S.; Panda, S.; Kim, H. J.; Sundaramoorthy, A.; Vivekananthan, V. Advances in Triboelectric Energy Harvesting at Liquid–Liquid Interfaces. *ACS Appl. Energy Mater.* **2025**, *8* (2), 659-682.
(6) Promsuwan, P.; Hasan, M. A. M.; Xu, S.; Yang, Y. Droplet nanogenerators: Mechanisms, performance, and applications. *Mater. Today.* **2024**, *80*, 497-528.
(7) Trung, V. D.; Le, P.-A.; Natsuki, J.; Zhao, W.; Phung, T. V. B.; Natsuki, T. Recent developments in droplet-based devices. *Mater. Today Chem.* **2024**, *36*, 101943.
(8) Wu, R.; Zhang, H.; Ma, H.; Zhao, B.; Li, W.; Chen, Y.; Liu, J.; Liang, J.; Qin, Q.; Qi, W.; et al. Synthesis, Modulation, and Application of Two-Dimensional TMD Heterostructures. *Chem. Rev.* **2024**, *124* (17), 10112-10191.
(9) Seol, M.; Kim, S.; Cho, Y.; Byun, K.-E.; Kim, H.; Kim, J.; Kim, S. K.; Kim, S.-W.; Shin, H.-J.; Park, S. Triboelectric Series of 2D Layered Materials. *Adv. Mater.* **2018**, *30* (39), 1801210.
(10) Ueberricke, L.; Coleman, J. N.; Backes, C. Robustness of Size Selection and Spectroscopic Size, Thickness and Monolayer Metrics of Liquid-Exfoliated WS2. *Phys. Status Solidi B* **2017**, *254* (11), 1700443.
(11) Pace, G.; del Rio Castillo, A. E.; Lamperti, A.; Lauciello, S.; Bonaccorso, F. 2D Materials-Based Electrochemical Triboelectric Nanogenerators. *Adv. Mater.* **2023**, *35* (23), 2211037.
(12) Synnatschke, K.; Müller, A.; Gabbett, C.; Mohn, M. J.; Kelly, A. G.; Mosina, K.; Wu, B.; Caffrey, E.; Cassidy, O.; Backes, C.; et al. Inert Liquid Exfoliation and Langmuir-Type Thin Film Deposition of Semimetallic Metal Diborides. *ACS Nano* **2024**, *18* (42), 28596-28608.
(13) Gabbett, C.; Doolan, L.; Synnatschke, K.; Gambini, L.; Coleman, E.; Kelly, A. G.; Liu, S.; Caffrey, E.; Munuera, J.; Murphy, C.; et al. Quantitative analysis of printed nanostructured networks using high-resolution 3D FIB-SEM nanotomography. **2024**, *15* (1), 278.
(14) Neilson, J.; Caffrey, E.; Cassidy, O.; Gabbett, C.; Synnatschke, K.; Schneider, E.; Munuera, J. M.; Carey, T.; Rimmer, M.; Sofer, Z.; et al. Production of Ultrathin and High-Quality Nanosheet Networks via Layer-by-Layer Assembly at Liquid–Liquid Interfaces. **2024**, *18* (47), 32589-32601.
(15) Synnatschke, K.; Cieslik, P. A.; Harvey, A.; Castellanos-Gomez, A.; Tian, T.; Shih, C.-J.; Chernikov, A.; Santos, E. J. G.; Coleman, J. N.; Backes, C. Length- and Thickness-Dependent Optical Response of Liquid-Exfoliated Transition Metal Dichalcogenides. *Chem. Mater.* **2019**, *31* (24), 10049-10062.
(16) Kubetschek, N.; Backes, C.; Goldie, S. Algorithm for Reproducible Analysis of Semiconducting 2D Nanomaterials Based on UV-VIS Spectroscopy. *Adv. Mater. Interfaces* **2024**, *11* (34), 2400311.
(17) Backes, C.; Paton, K. R.; Hanlon, D.; Yuan, S.; Katsnelson, M. I.; Houston, J.; Smith, R. J.; McCloskey, D.; Donegan, J. F.; Coleman, J. N. Spectroscopic metrics allow in situ measurement of mean size and thickness of liquid-exfoliated few-layer graphene nanosheets. *Nanoscale* **2016**, *8* (7), 4311-4323, 10.1039/C5NR08047A.
(18) Backes, C.; Campi, D.; Szydlowska, B. M.; Synnatschke, K.; Ojala, E.; Rashvand, F.; Harvey, A.; Griffin, A.; Sofer, Z.; Marzari, N.; et al. Equipartition of Energy Defines the Size–Thickness Relationship in Liquid-Exfoliated Nanosheets. **2019**, *13* (6), 7050-7061.
(19) Liu, S.; Xu, W.; Yang, J.; Liu, S.; Meng, Y.; Jia, L.; Chen, G.; Qin, Y.; Li, X. DC Output Water Droplet Energy Harvester Enhanced by the Triboelectric Effect. *ACS Appl. Electron. Mater.* **2022**, *4* (6), 2851-2858.
(20) Liu, X.; Yu, A.; Qin, A.; Zhai, J. Highly Integrated Triboelectric Nanogenerator for Efficiently Harvesting Raindrop Energy. **2019**, *4* (11), 1900608.

(21) Hu, Y.; Sun, R.; Li, S.; Liu, C.; Zhao, J.; Mo, J.; Luo, D.; Pan, Y. Optimizing raindrop energy harvesting: Exploring water droplet spreading effects on IDE-based TENG for sustainable power generation. **2024**, *123*, 109358.
(22) Wu, X.; Cai, T.; Wu, Q.; Meng, J.; Wang, W.; Li, W.; Hu, C.; Zhang, X.; Wang, D. Droplet-Based Triboelectric Nanogenerators with Needle Electrodes for Efficient Water Energy Harvesting. **2025**, *17* (9), 13762-13772.
(23) Yang, C.; Wang, Y.; Wang, Y.; Wang, Z.; Guo, Y.; Zhang, L.; Liu, X.; Chen, H. Skeleton Enhanced Dispersed Lubricant Particle Based Triboelectric Nanogenerator for Droplet Energy Harvesting. **2025**, *n/a* (n/a), e05363.
(24) Kim, H.-g.; Choi, H. J. Thickness dependence of work function, ionization energy, and electron affinity of Mo and W dichalcogenides from DFT and GW calculations. **2021**, *103* (8), 085404.
(25) Wang, J.; Xu, S.; Hu, C. Charge Generation and Enhancement of Key Components of Triboelectric Nanogenerators: A Review. *Adv. Mater.* **2024**, *36* (50), 2409833.
(26) Laturia, A.; Van de Put, M. L.; Vandenberghe, W. G. Dielectric properties of hexagonal boron nitride and transition metal dichalcogenides: from monolayer to bulk. **2018**, *2* (1), 6.
(27) Karger, L.; Synnatschke, K.; Settele, S.; Hofstetter, Y. J.; Nowack, T.; Zaumseil, J.; Vaynzof, Y.; Backes, C. The Role of Additives in Suppressing the Degradation of Liquid-Exfoliated WS2 Monolayers. **2021**, *33* (42), 2102883.
(28) Nguyen, T. V.; Tekalgne, M.; Tran, C. V.; Nguyen, T. P.; Dao, V.; Le, Q. V.; Ahn, S. H.; Kim, S. Y. Synthesis of MoS2/WS2 Nanoflower Heterostructures for Hydrogen Evolution Reaction. **2024**, *2024* (1), 3192642.
(29) Miakota, D. I.; Unocic, R. R.; Bertoldo, F.; Ghimire, G.; Engberg, S.; Geohegan, D.; Thygesen, K. S.; Canulescu, S. A facile strategy for the growth of high-quality tungsten disulfide crystals mediated by oxygen-deficient oxide precursors. **2022**, *14* (26), 9485-9497, 10.1039/D2NR01863B.
(30) Sasikala, S. P.; Prabhakaran, P.; Baskaran, S.; Kim, J. T.; Lee, G. S.; Yoon, Y. H.; Choi, H. J.; Kim, J. G.; Kim, J. B.; Kim, S. O. Direct Solution-Phase Synthesis and Functionalization of 2D WSe2 for Ambient Stability. **2023**, *29* (61), e202301744.
(31) Gabbett, C.; Kelly, A. G.; Coleman, E.; Doolan, L.; Carey, T.; Synnatschke, K.; Liu, S.; Dawson, A.; O'Suilleabhain, D.; Munuera, J.; et al. Understanding how junction resistances impact the conduction mechanism in nano-networks. *Nat. Commun.* **2024**, *15* (1), 4517.
(32) Xie, Y.; Zheng, J.; Guo, J.; Huang, H.; Lin, W.; Liao, J.; Guo, Q.; Duan, J.; Tang, Q.; Yang, X. Triboelectricity-enhanced photovoltaic effect in hybrid tandem solar cell under rainy condition. *Nano Energy* **2025**, *135*, 110647.
(33) Gabbett, C.; Boland, C. S.; Harvey, A.; Vega-Mayoral, V.; Young, R. J.; Coleman, J. N. The Effect of Network Formation on the Mechanical Properties of 1D:2D Nano:Nano Composites. *Chem. Mater.* **2018**, *30* (15), 5245-5255.
(34) Ye, C.; Liu, D.; Chen, P.; Cao, L. N. Y.; Li, X.; Jiang, T.; Wang, Z. L. An Integrated Solar Panel with a Triboelectric Nanogenerator Array for Synergistic Harvesting of Raindrop and Solar Energy. *Adv. Mater.* **2023**, *35* (11), 2209713.

# Supporting information

# Toward All 2D-based Printed Raindrop Triboelectric Nanogenerators

*Foad Ghasemi[1*], Jonas Heirich[1], Dimitri Sharikow[1], Sebastian Klenk[2], Jonathan N. Coleman[3], Georg S. Duesberg[2], Claudia Backes[1*]*

4.  Physical Chemistry of Nanomaterials and CINSaT, Kassel University, Heinrich-Plett-Str. 40, 34132, Kassel, Germany.

5.  Institute of Physics, University of the Bundeswehr Munich, Werner-Heisenberg-Weg 39 85577 Neubiberg, Germany.

6.  School of Physics and CRANN, Trinity College, Dublin 2, Ireland

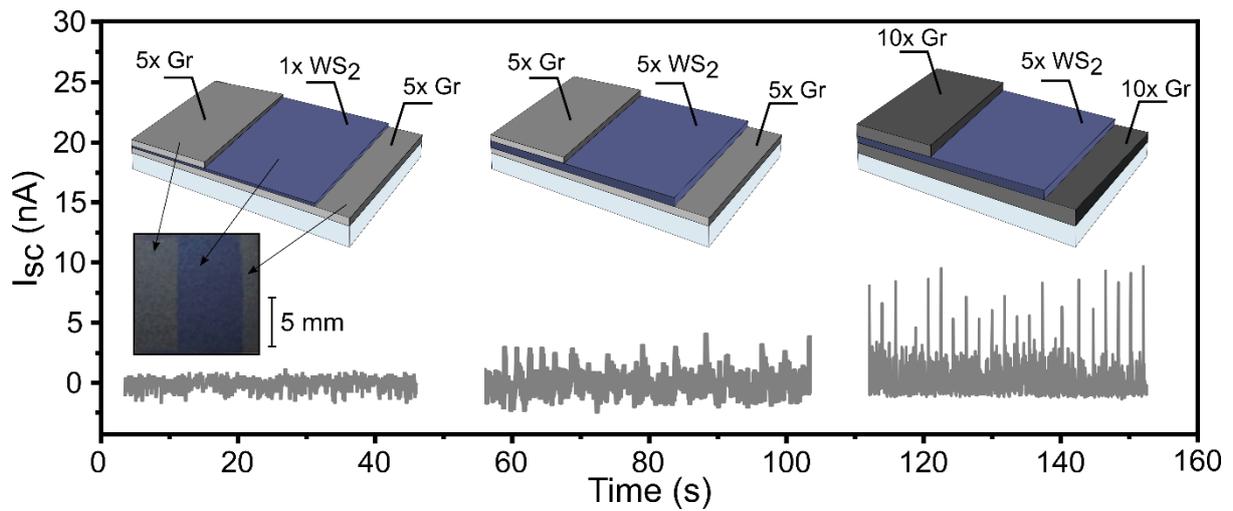

Figure S1. Proof-of-concept RD-TENG device geometries tested using LPE-derived TMD films. **Left**: The 5x bottom graphene, 1x WS$_2$, 5x top graphene, **middle**: The 5x bottom graphene, 5x WS$_2$, 5x top graphene, **right**: The 10x bottom graphene, 5x WS$_2$, 10x top graphene 2D based RD-TENGs. Increasing the graphene deposition cycles resulted in stronger short-circuit current spikes due to improved electrical conductivity of the contacts.

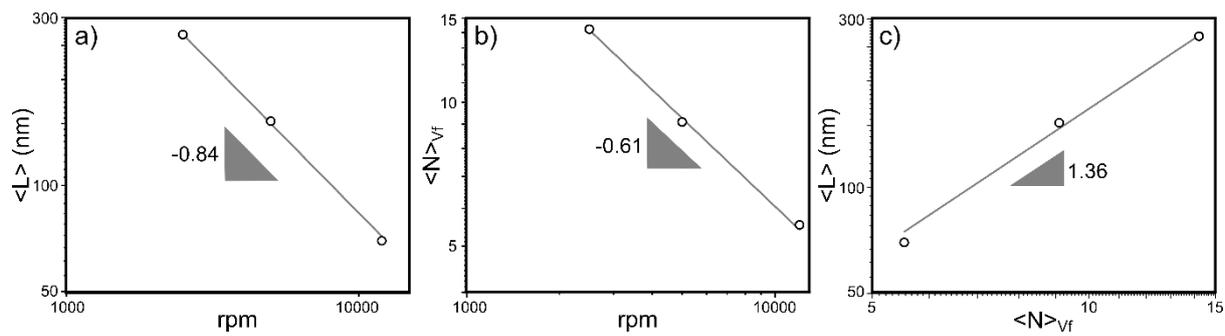

Figure S2. WS$_2$ average (a) length and (b) volume-fraction weighted number of layers as a function of centrifuge speed. (c) average WS$_2$ flake size <L> vs number of layers <N>$_{Vf}$.

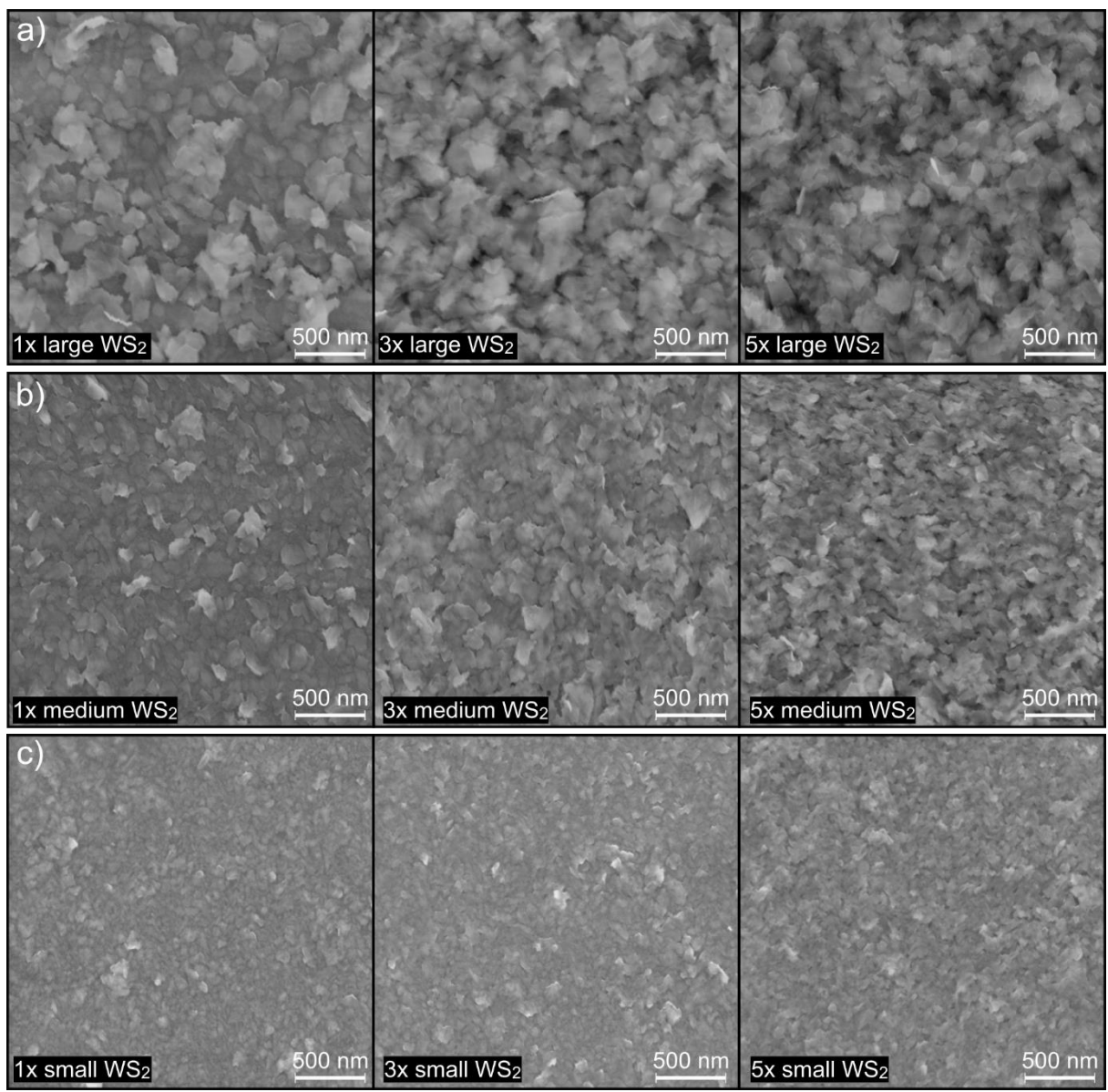

Figure S3. SEM images of WS$_2$ films for a) large, b) medium, and c) small flakes after 1x, 3x, and 5x deposition cycles.

**Surface coverage calculation:**

Surface coverage was estimated from SEM images for each fraction after 1 to 5 deposition cycles. Gwyddion software was used for image processing, following a standardized procedure. First, the "Mark by Threshold" function under Data Process/Grains was applied, and parameters such as height, rank, slope, and curvature were adjusted to accurately identify and colorize the WS$_2$ flakes while ensuring the gaps between them remained visible. Once the flake regions were distinguished, the "Grain Summary function" was used to extract the total projected area, which was recorded as the surface coverage.

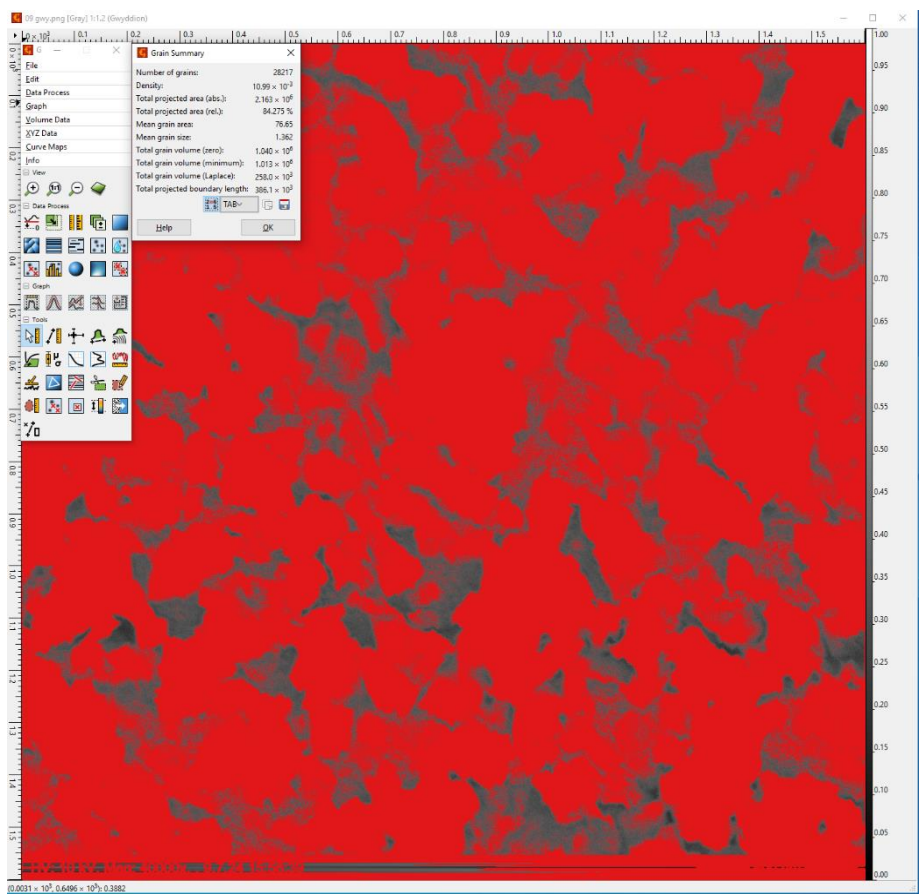

Figure S4. Calculation of surface coverage with Gwydion software. The deposited flakes are highlighted in red to visualize coverage across the substrate surface.

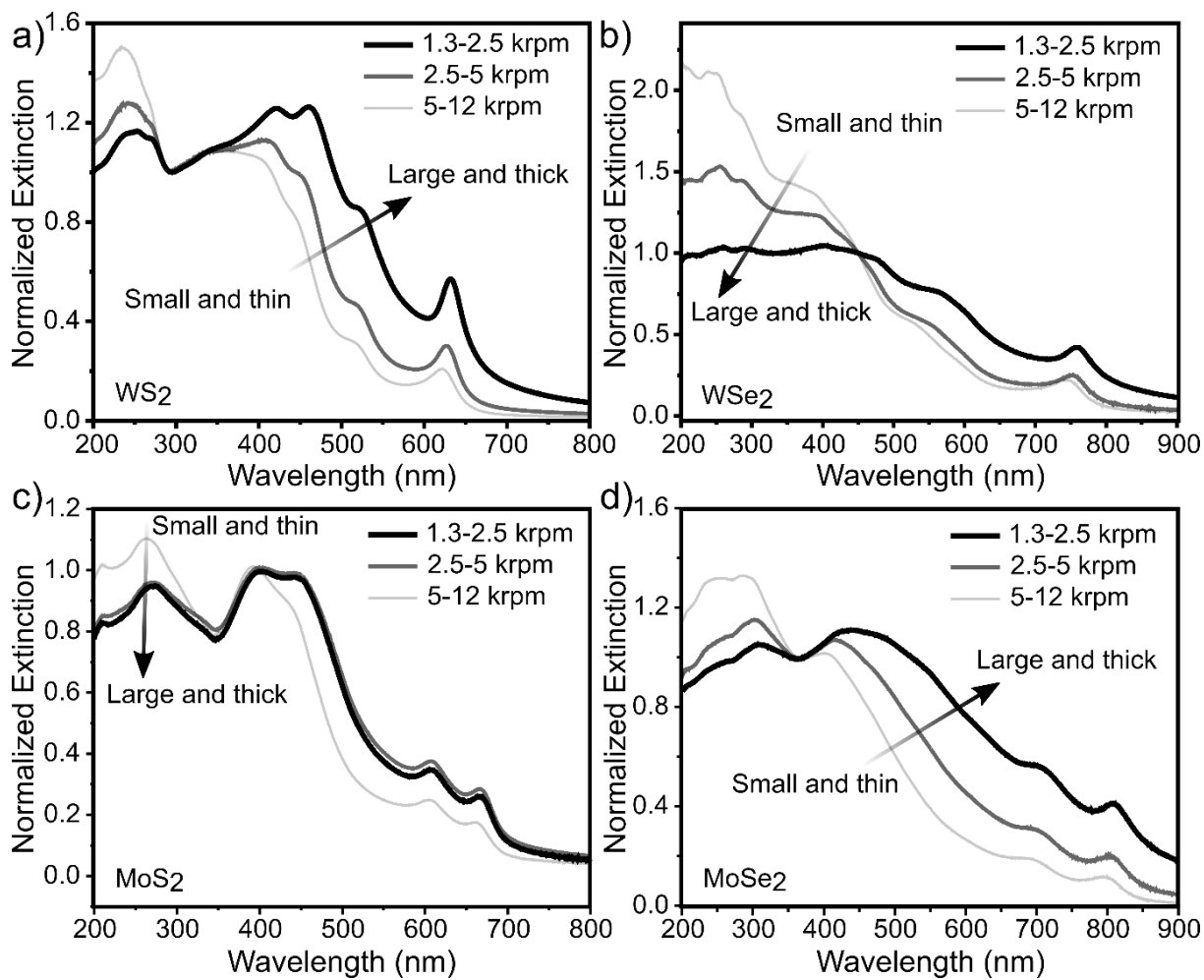

Figure S5. Normalized extinction spectra of the exfoliated (a) WS$_2$, (b) WSe$_2$, (c) MoS$_2$, and (d) MoSe$_2$ fractions with different sizes and thicknesses.

Table S1. Flake's size and thickness of the medium-sized TMD fraction calculated from the extinction spectra according to ref 1

| TMD | <L> | <N>$_{vf}$ |
|---|---|---|
| WS$_2$ | 152.24 | 9.09 |
| MoS$_2$ | 167.75 | 8.66 |
| WSe$_2$ | 158.61 | 8.19 |
| MoSe$_2$ | 150.65 | 7.71 |

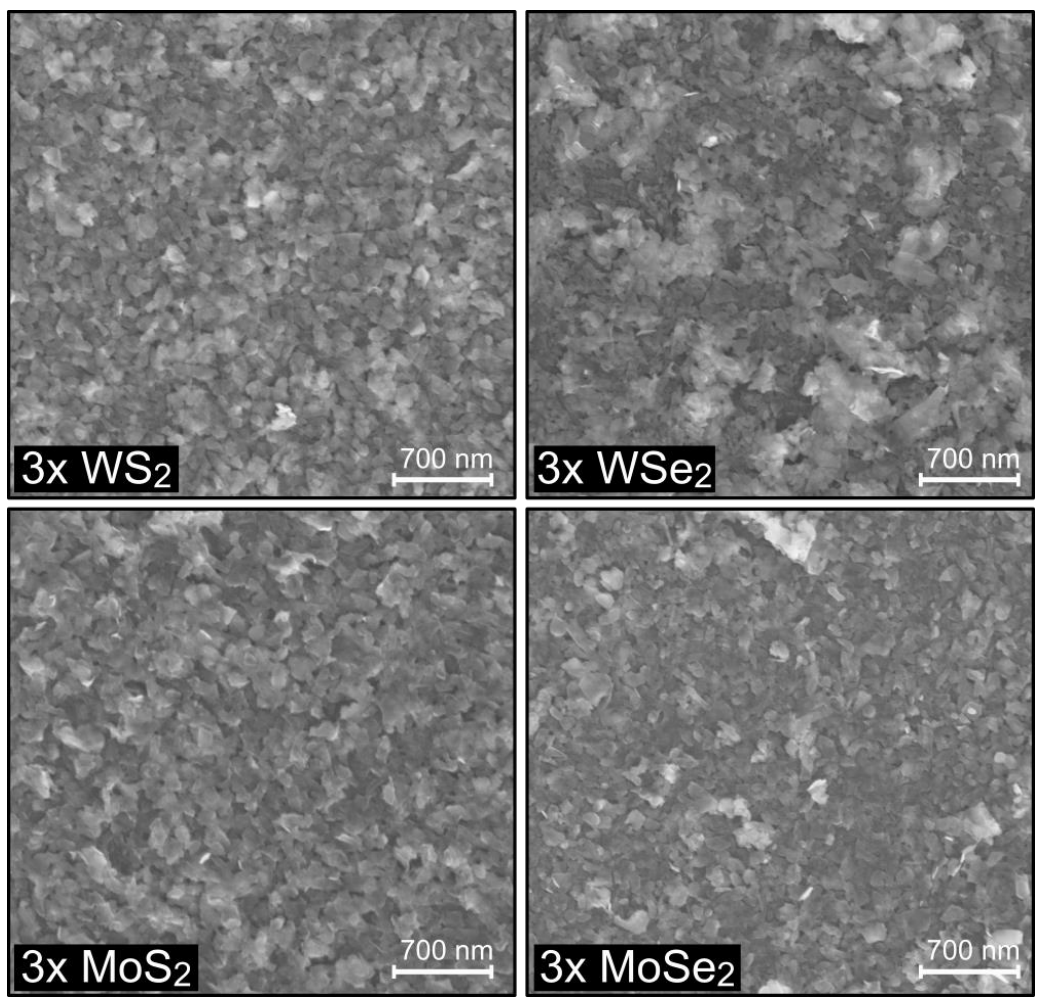

Figure S6. SEM images of the TMD films on ITO substrates with 3x deposition cycles.

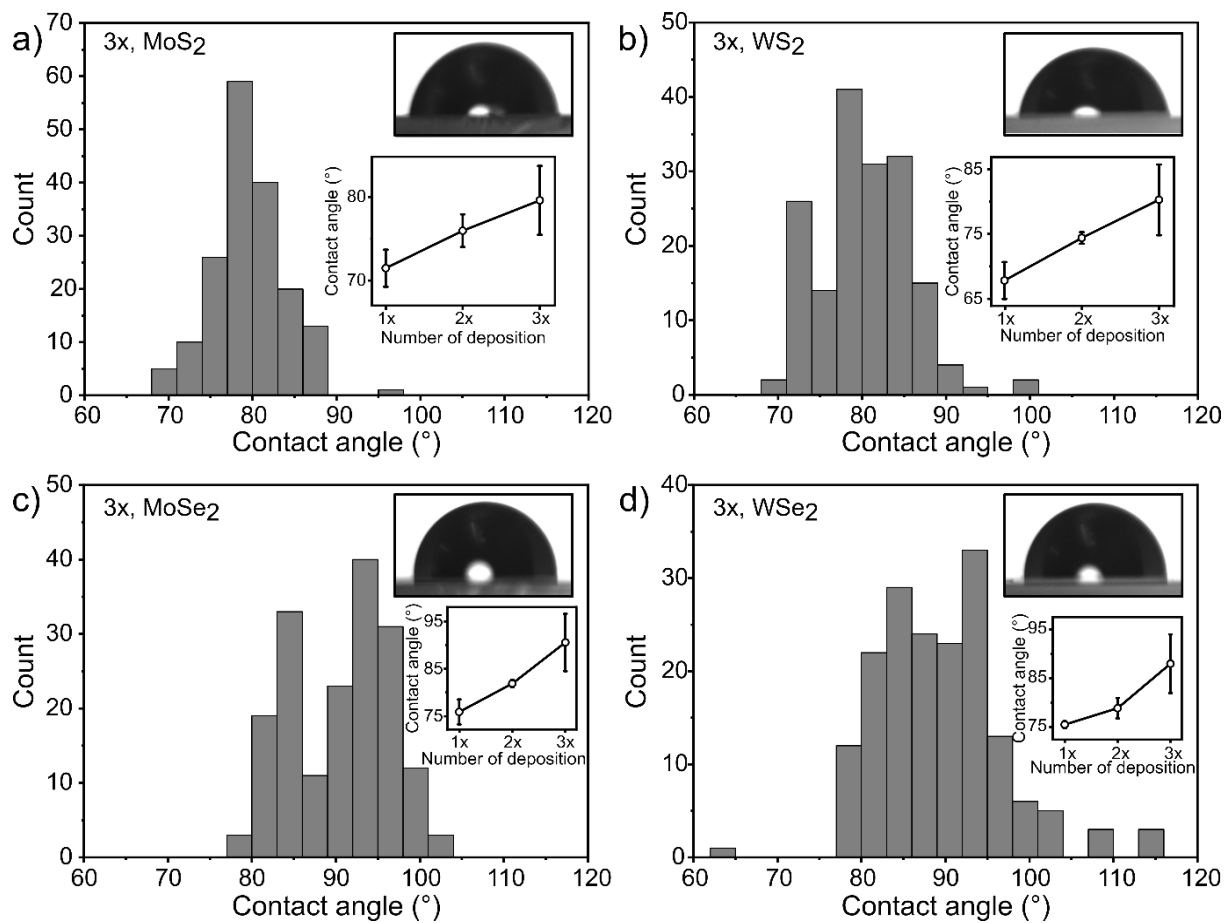

Figure S7. Contact angle histograms of the 3x deposited medium-sized (a) $MoS_2$, (b) $WS_2$, (c) $MoSe_2$, and (d) $WSe_2$ films on silicon substrates. The optical images of the DI drop released onto the films and contact angles for 1x to 3x films are presented in the inset.

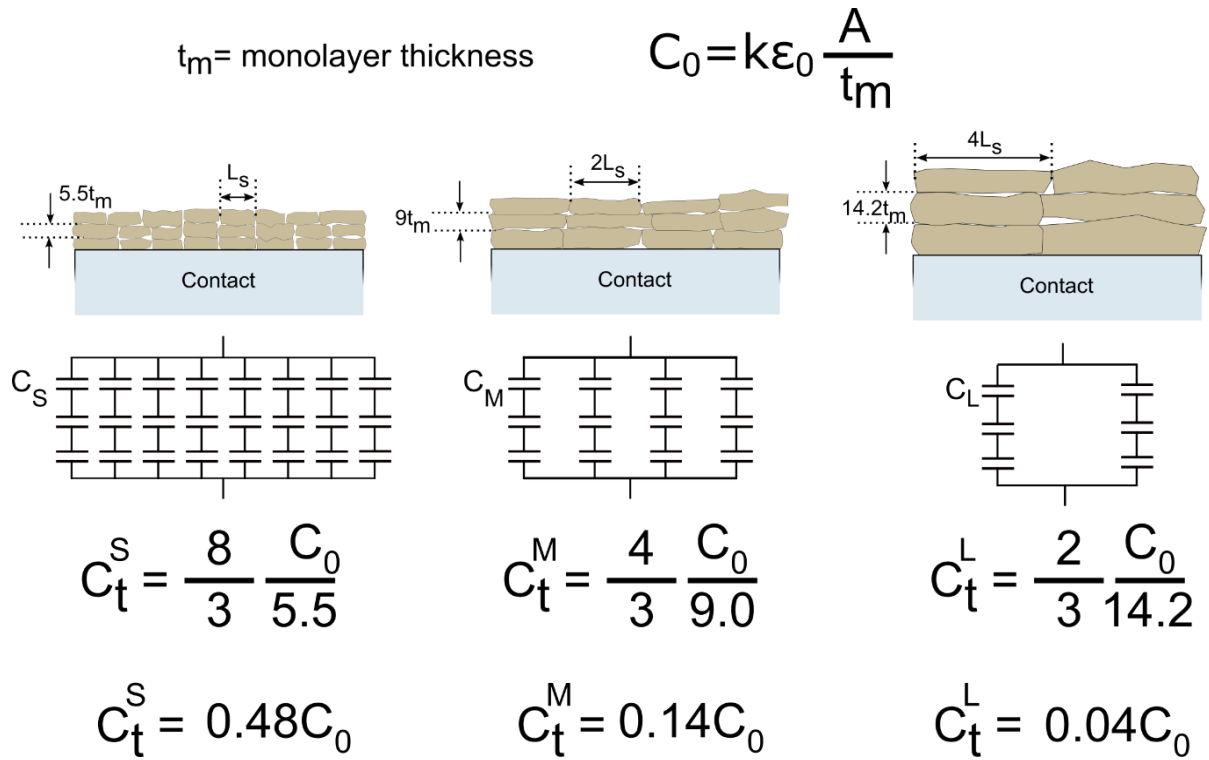

Figure S8. Schematic illustration and equivalent circuit models of films composed of $WS_2$ flakes with different sizes. Each flake is modeled as a capacitor with capacitance inversely proportional to its thickness ($C = C_0/N$, where $C_0$ is the capacitance of a monolayer and N is the average number of layers). To cover a given area, a larger number of small flakes is required compared to medium and large flakes, resulting in more capacitors in parallel and, therefore, larger total capacitance. The equivalent total capacitances for films composed of small, medium, and large flakes are approximately $0.48C_0$, $0.14C_0$, and $0.04C_0$, respectively.

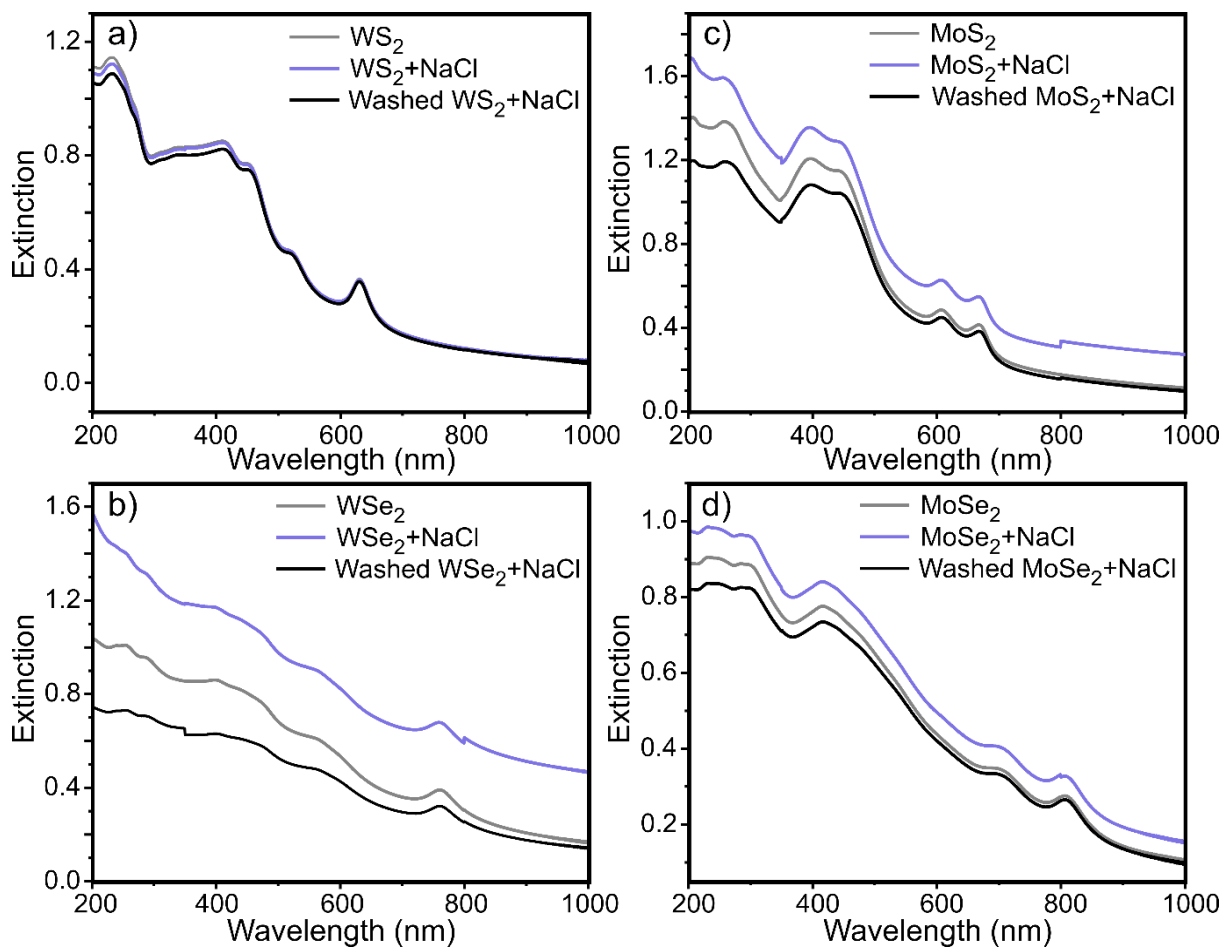

Figure S9. UV-Vis extinction spectra of freshly prepared (a) $WS_2$, (b) $WSe_2$, (c) $MoS_2$, and (d) $MoSe_2$ films, followed by spectra recorded after impact of NaCl droplets and subsequent rinsing with DI water.

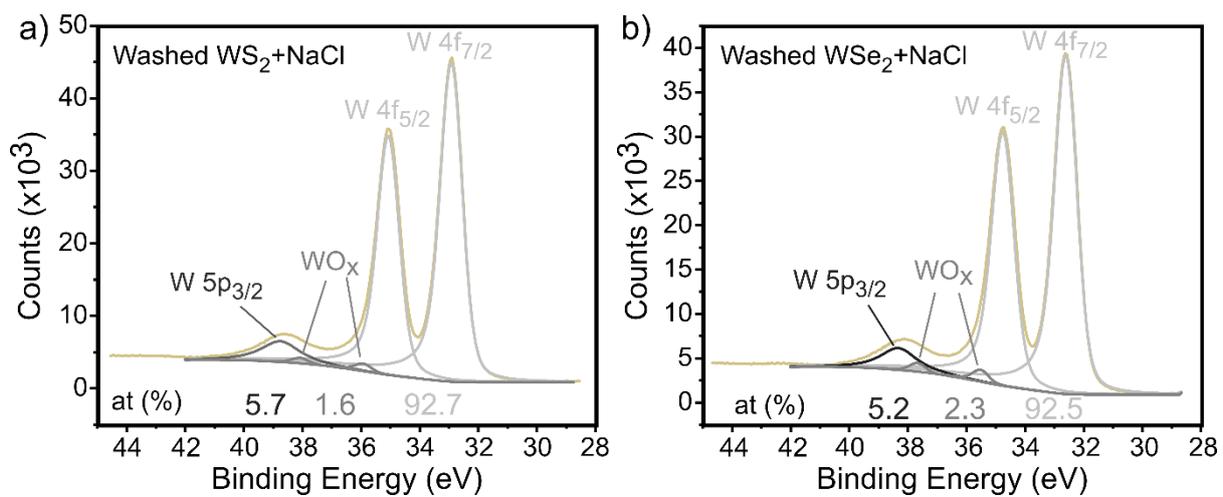

Figure S10. XPS spectra of W 4f core level for NaCl-affected (a) $WS_2$, and (b) $WSe_2$ films after washing with DI water.

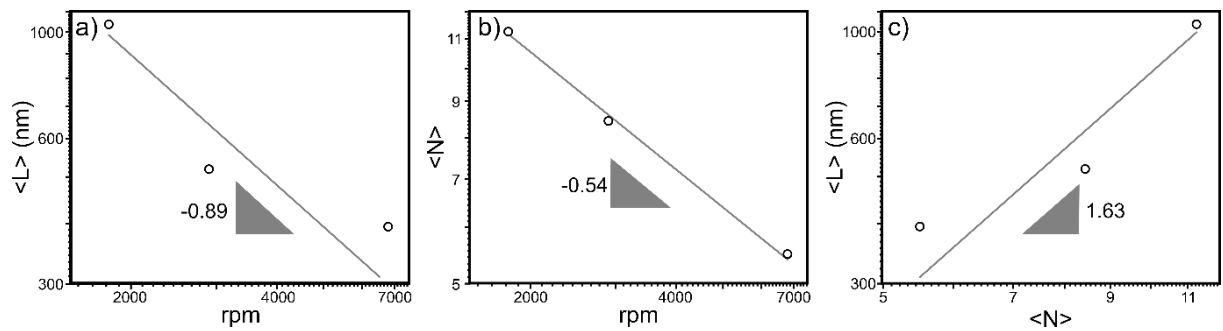

Figure S11. Graphene (a) average length and (b) arithmetic average number of layers as a function of centrifuge speed. (c) average flake size <L> vs number of layers <N>.

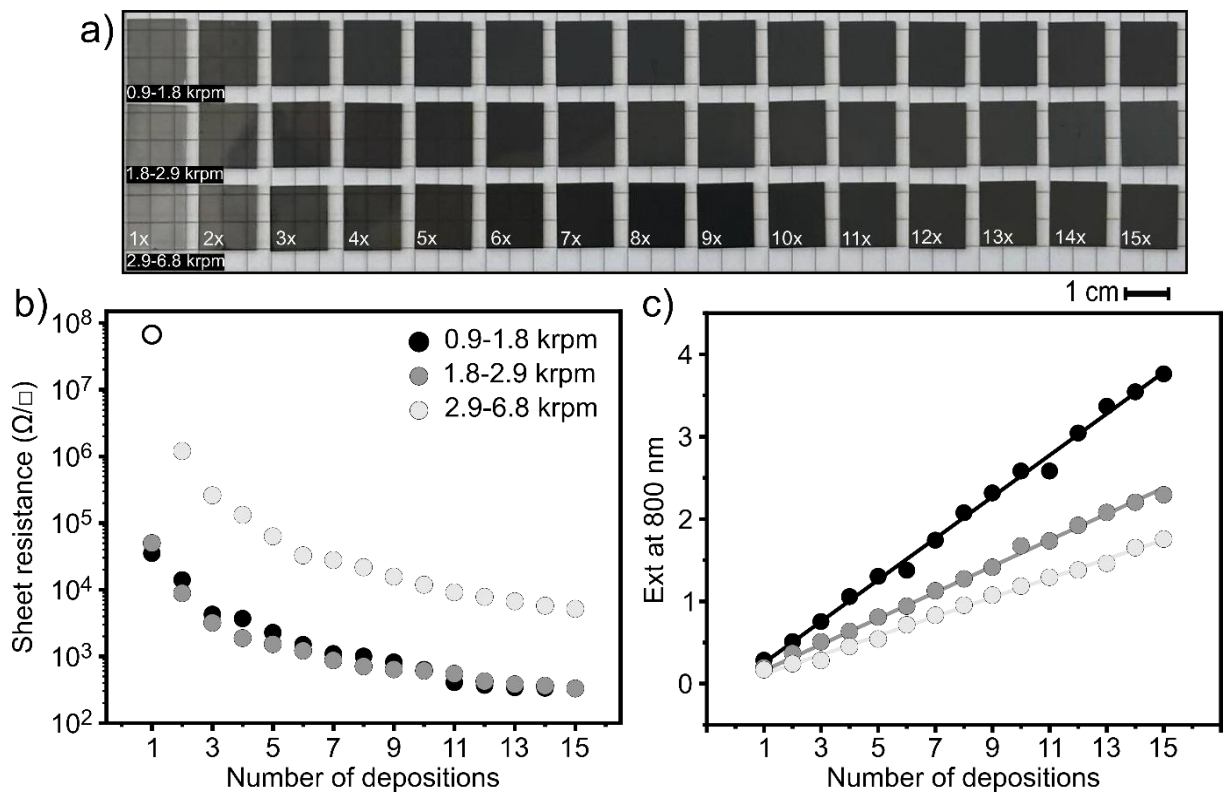

Figure S12. a) Optical images of graphene films for 1x to 15x deposition cycles. b) Sheet resistance, and c) extinction values at 800 nm of graphene films at various deposition cycles (1x to 15x),

Table S2. The sheet resistance of the graphene films as a function of deposition cycles.

| Deposition cycles | 0.9-1.8 krpm (kΩ/□) | 1.8-2.9 krpm (kΩ/□) | 2.9-6.8 krpm (kΩ/□) |
|---|---|---|---|
| 1x | 35.16 | 50.07 | 66847.00 |
| 2x | 14.00 | 8.83 | 1200.98 |
| 3x | 4.26 | 3.17 | 260.59 |
| 4x | 3.67 | 1.87 | 131.69 |
| 5x | 2.26 | 1.51 | 62.90 |
| 6x | 1.49 | 1.21 | 32.49 |
| 7x | 1.08 | 0.87 | 27.64 |
| 8x | 0.99 | 0.70 | 21.66 |
| 9x | 0.81 | 0.63 | 15.68 |
| 10x | 0.62 | 0.60 | 11.78 |
| 11x | 0.40 | 0.54 | 9.10 |
| 12x | 0.36 | 0.42 | 7.79 |
| 13x | 0.34 | 0.38 | 6.66 |
| 14x | 0.33 | 0.36 | 5.71 |
| 15x | 0.32 | 0.32 | 5.16 |

**Calculation of mass ratio in Graphene/SWCNT dispersion:**

To calculate the mass ratio of graphene to carbon nanotubes (SWCNTs), first, SWCNT dispersion was prepared in IPA/DI water and filtered using a Cytiva Whatman filter (0.02 μm pore size) to determine the mass of dispersed nanotubes and thus to calculate the concentration of the dispersion and the extinction coefficient. The extinction coefficient of the SWCNT dispersion was calculated to be 3292 mL.mg$^{-1}$.m$^{-1}$ based on UV-Vis extinction measurements and gravimetric concentration. Then, different volumes of the SWCNT dispersion were mixed with the graphene dispersion. Using the formula:

$$SWCNT\ wt\% = \left(\frac{mass\ of\ SWCNTs\ per\ volume}{total\ mass\ (SWCNTs + graphene)\ per\ volume}\right) * 100$$

The final SWCNT weight percentage was calculated.

Table S3. The sheet resistance of the graphene films with the addition of varying ratios of SWCNTs.

| Deposition cycles | 0.9-1.8 krpm Gr (Ω/□) | Gr:CNT 8% (Ω/□) | Gr:CNT 27% (Ω/□) | Gr:CNT 60% (Ω/□) | Gr:CNT 75% (Ω/□) |
|---|---|---|---|---|---|
| **1x** | 50078.6 | 8869.1 | 2660.2 | 2093.7 | 1572.6 |
| **2x** | 8837.4 | 1844.5 | 906.4 | 697.9 | 641.9 |
| **3x** | 3172.4 | 821.6 | 551.9 | 404.9 | 261.5 |
| **4x** | 1876.2 | 581.4 | 352.1 | 282.7 | 192.5 |
| **5x** | 1518.2 | 492.1 | 277.5 | 204.1 | 189.9 |

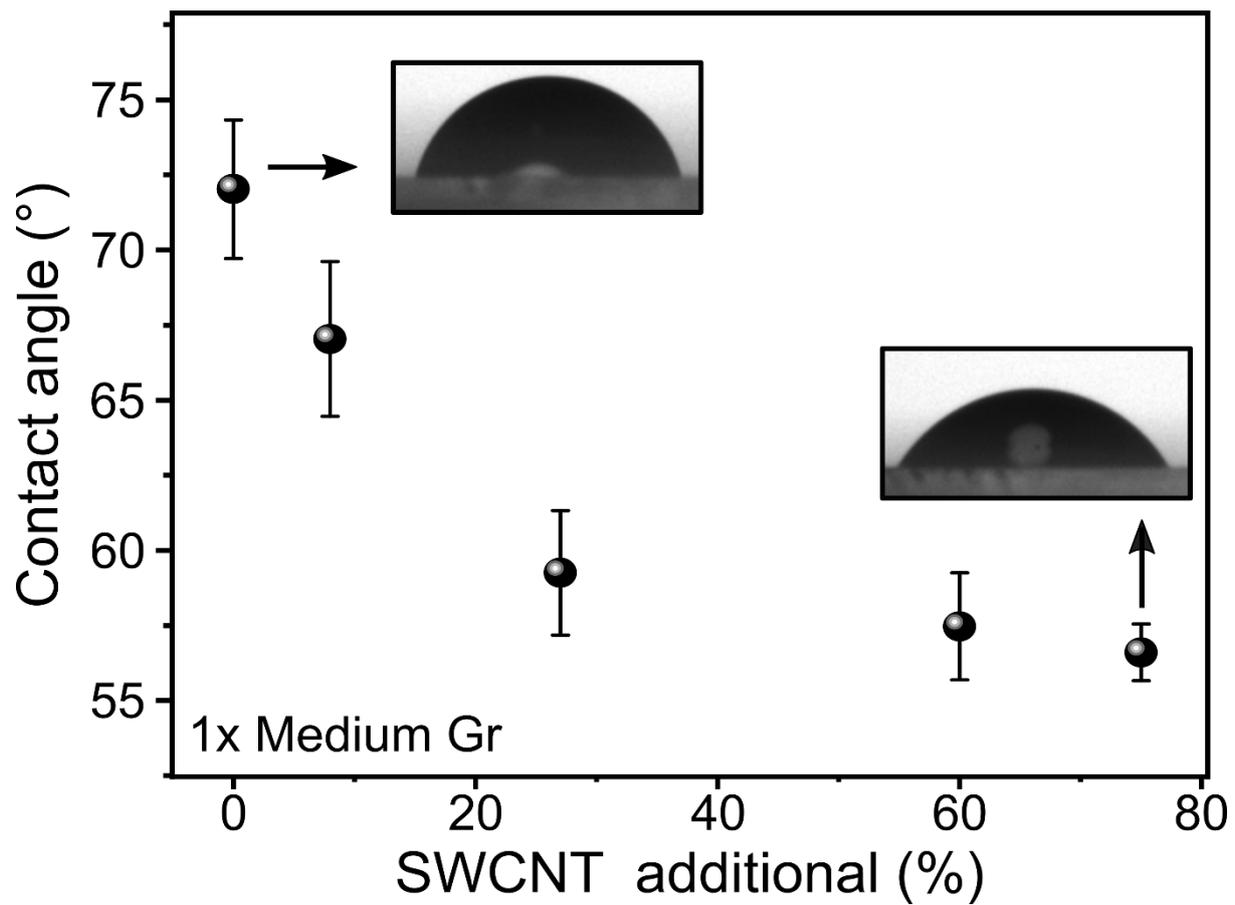

Figure S13. Contact angle measurement of the 1x medium-sized graphene film with different mass ratios of SWCNTs.

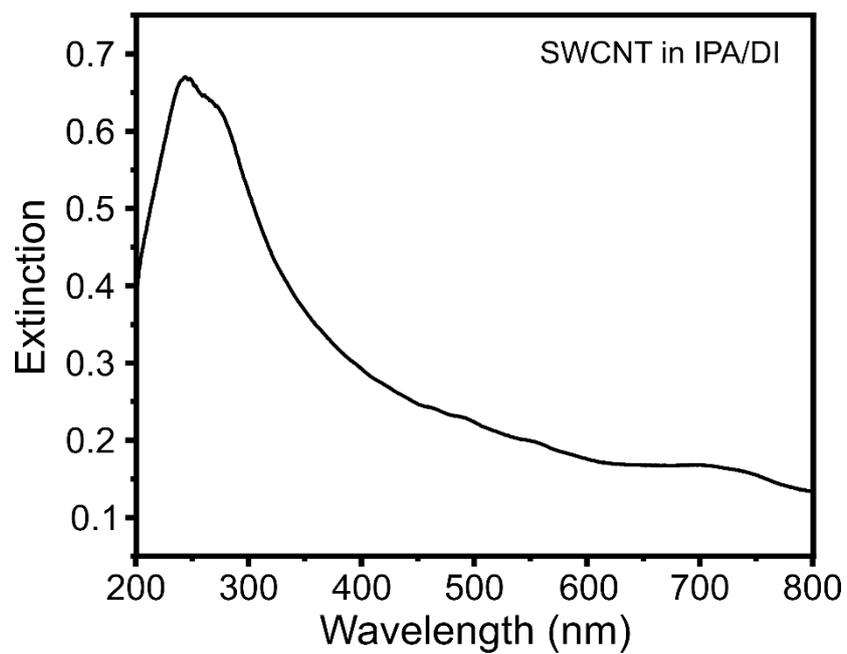

Figure S14. Extinction spectrum of the exfoliated SWCNTs in the IPA/DI solution.

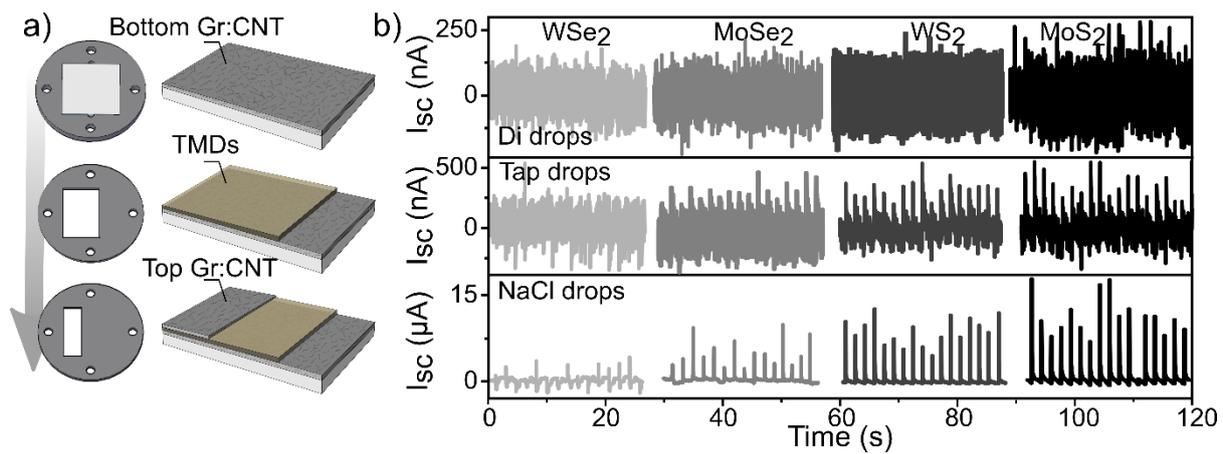

Figure S15. (a) Fabrication steps for vertically structured 2D-based RD-TENGs. (b) Short-circuit current ($I_{sc}$) responses to DI water, tap water, and 1 M NaCl drops.

1   Kubetschek, N., Backes, C. & Goldie, S. Algorithm for Reproducible Analysis of Semiconducting 2D Nanomaterials Based on UV-VIS Spectroscopy. *Advanced Materials Interfaces* **11**, 2400311 (2024). https://doi.org:https://doi.org/10.1002/admi.202400311